\documentclass[aps,prd,twocolumn,amsmath,amssymb,superscriptaddress,showkeys,showpacs,floatfix]{revtex4-1}
\usepackage[final]{graphicx}
\usepackage{hyperref}
\usepackage{amsmath}
\usepackage{bbm}
\usepackage{amsfonts}
\usepackage{amssymb}
\usepackage{latexsym}
\usepackage{graphicx}
\usepackage[english]{babel}
\usepackage{multirow}
\usepackage{float}
\usepackage{url}
\usepackage{slashed}
\usepackage{xcolor}
\usepackage[utf8]{inputenc}
\usepackage{verbatim}
\usepackage{lipsum}

\usepackage[utf8]{inputenc}

\usepackage{diagbox}

\usepackage{mathtools}
\usepackage{amsfonts}
\usepackage{mathrsfs}
\usepackage{bbm}
\usepackage{slashed}

\usepackage{graphicx}
\usepackage{color}
\usepackage{array}
\usepackage{esint}
\usepackage{placeins}
\usepackage{booktabs}
\usepackage{makecell}
\usepackage{epstopdf}
\usepackage[caption=false]{subfig}

\usepackage{xspace}
\usepackage{siunitx}
\usepackage{hyperref}
\usepackage[nameinlink]{cleveref}
\usepackage{appendix}

\usepackage{xifthen}
\usepackage{xcolor}
\hypersetup{
	colorlinks,
	linkcolor={red!75!black},
	citecolor={blue!75!black},
	urlcolor={blue!75!black}
}
\usepackage{comment}

\newcommand{\be}{\begin{equation}}
\newcommand{\ee}{\end{equation}}
\newcommand{\ba}{\begin{array}}
\newcommand{\ea}{\end{array}}

\newcommand{\Tr}{{\rm tr}}

\usepackage{ulem}
\newcommand{\beq}{\begin{eqnarray}}
\newcommand{\eeq}{\end{eqnarray}}

\newcommand{\ii}{\mathrm{i}}

\begin{document}

\title{Impact of Primordial Magnetic Fields on the First-Order Electroweak Phase Transition}
\author{Yuefeng Di}
\affiliation{
Institute of Theoretical Physics, 
Chinese Academy of Sciences, Beijing 100190, China}
\affiliation{School of Physical Sciences, University of Chinese Academy of Sciences (UCAS), Beijing 100049, China}

\author{Ligong Bian\thanks{Corresponding Author.}}
\email{lgbycl@cqu.edu.cn}
\affiliation{Department of Physics and Chongqing Key Laboratory for Strongly Coupled Physics, Chongqing University, Chongqing 401331, P. R. China}
\author{Rong-Gen Cai}
\email{cairg@itp.ac.cn}
\affiliation{
Institute of Fundamental Physics and Quantum Technology, Ningbo University, Ningbo, 315211, China}

\date{\today}

\begin{abstract}
We numerically study how the primordial‌ magnetic field affects the first-order electroweak phase transition in the early Universe. We observe that: 1) the phase transition process would be slowed down by the magnetic field; 2) the phenomenon of vortex structure of the Higgs condensation appears when the homogenesis hypermagentic field $g' B_Y^{ex}/m_W^2\gtrsim3.63$; and, 3) the helical hypermagnetic field can dramatically enhance the sphaleron rate and validate the generation of the baryon asymmetry through the chiral anomaly. 

\end{abstract}
\maketitle

\section{Introduction}
 Recently, the evidence of the stochastic gravitational wave background (SGWB) has been reported by pulsar timing array experiments around the world, PPTA
~\cite{Reardon:2023gzh}, EPTA and InPTA~\cite{EPTA:2023fyk}, NanoGrav~\cite{NANOGrav:2023gor}, and CPTA~\cite{Xu:2023wog}.
The first-order PT
at QCD scale has been considered as one compelling possibility to address the  signatures~\cite{Bian:2023dnv,Ghosh:2023aum,He:2023ado,Zheng:2024tib,Winkler:2024olr,Chen:2023bms,Bringmann:2023opz}. The SGWB produced by the strong first-order electroweak PT~\cite{Mazumdar_2019, Caprini_2016, Caprini_2020} might be able to detect by future space-based interferometer, such as the Laser Interferometer Space Antenna (LISA)~\cite{LISA:2017pwj},  Taiji~\cite{taiji}, and TianQin~\cite{Luo_2016}. Aside from SGWB, during the first-order electroweak PT, it was found that the bubble collision dynamics can drive the production of the primordial magnetic fields (MFs) ~\cite{PhysRevLett.51.1488, Di_2021, Enqvist:1993np,Yang_2022,Vachaspati:1991nm,Baym:1995fk,Grasso:1997nx}. The MFs generated in the early Universe
might influence the Higgs vacuum and yield formation of a typical vortex structure known as the Ambj\o rn-Olesen condensation~\cite{Nielsen:1978rm, PhysRevD.45.3833, SALAM1975203, LINDE1976435,Ambjorn_1989, Chernodub_2023}. MFs might affect the electroweak sphaleron rate during the electroweak cross-over~\cite{Annala:2023jvr,Ho:2020ltr} and the first-order PT~\cite{Comelli:1999gt}.
After considering amplification mechanisms~\cite{Grasso_2001,Widrow_2002,Kulsrud_2008,Kandus_2011, Widrow_2011,Durrer_2013}, these MFs can explain the MFs in the Milky Way~\cite{Wielebinski2005}, galaxy clusters~\cite{Clarke_2001, Bonafede_2010, Feretti_2012}, and the voids of large-scale structures~\cite{Dolag_2010, Tavecchio_2010, Tavecchio_2011, Vovk_2012}, and be probed
by gamma-ray observations of blazars~\cite{Dermer_2011, Taylor_2011, Neronov_2010}.

The first-order electroweak PT, as a generic prediction of new physics beyond the Standard Model (SM) of particle physics~\cite{Hindmarsh_2021, Caldwell:2022qsj, Athron_2024}, provides one of the three Sakharov conditions~\cite{Sakharov:1967dj}, i.e., the out-of-equilibrium conditions, and makes it possible to explain the baryon asymmetry of the Universe through electroweak baryogenesis~\cite{Morrissey_2012, Cohen_2012}. 
Different from the microphysics for baryogenesis, Refs.~\cite{Giovannini_1998, Giovannini:1997eg, Joyce:1997uy, Fujita:2016igl} build the relationship between the baryon number and the helicity of primordial MF through the abelian anomaly.
Refs.~\cite{Kamada:2016eeb, Kamada:2016cnb} further found that source terms
arising from the chiral anomaly in the background of a helical hypermagnetic field
can drive the baryon asymmetry generation after considering the decay of MF helicity around the SM cross-over.

In this work, we intend to study the first-order electroweak PT in a MF background. We conduct lattice field simulations to include the non-linear dynamics during the PT. We analyze the impact of the ‌primordial‌ MFs on the PT speed and investigate the Ambj\o rn-Oleson condensation during the PT. We also evaluate how the helical magnetic field affects the Chern-Simons number evolution, altering the sphaleron rate. Finally, we observe that the helical MF can alter the baryon number and the lepton number during first-order PT, which provides a chance to address the baryon asymmetry of the Universe.

\section{The physical picture}

We consider an external $\mathrm{U(1)_Y}$ gauge field $Y_\mathrm{ex}^\mu$ that does not vary over time to achieve an external MF, and the complete electroweak Lagrangian is  
\begin{align}
    \mathcal{L} =\ &(D_\mu\Phi)^\dagger(D^\mu\Phi)
    - \frac14 W^a_{\mu\nu}W^{a\mu\nu}
    - \frac14 Y_{\mu\nu}Y^{\mu\nu}\notag\\ 
    &\ - \frac{1}{2} Y^{\mathrm{ex}}_{\mu\nu} Y^{\mu\nu} - \mathcal{V} (\Phi)\;.\label{L}
\end{align}
$\Phi$ is the Higgs doublet, and $W^a_{\mu\nu}$ and $Y_{\mu\nu}$ represent the field strengths of the $\mathrm{SU(2)_L}$ and $\mathrm{U(1)_Y}$ gauge fields, respectively. 
Here we omit $Y^\mathrm{ex}_{\mu\nu} Y^{\mu\nu}_{\mathrm{ex}}$ since it contributes only a constant. 
$D_\mu$ is the covariant derivative:
\begin{align}
    D_\mu = \partial_\mu - \ii g\frac{\sigma^a}{2} W_\mu^a - 
    \ii g' \,\frac12(Y_\mu + Y^\mathrm{ex}_\mu), \label{D}
\end{align}
where $\sigma^a,\ a=1,\ 2,\ 3$ are Pauli matrices, and the coupling constants are $g = 0.65$ and $g' = 0.53g$. 

\begin{figure}
    \centering
    \includegraphics[width=0.4\textwidth]{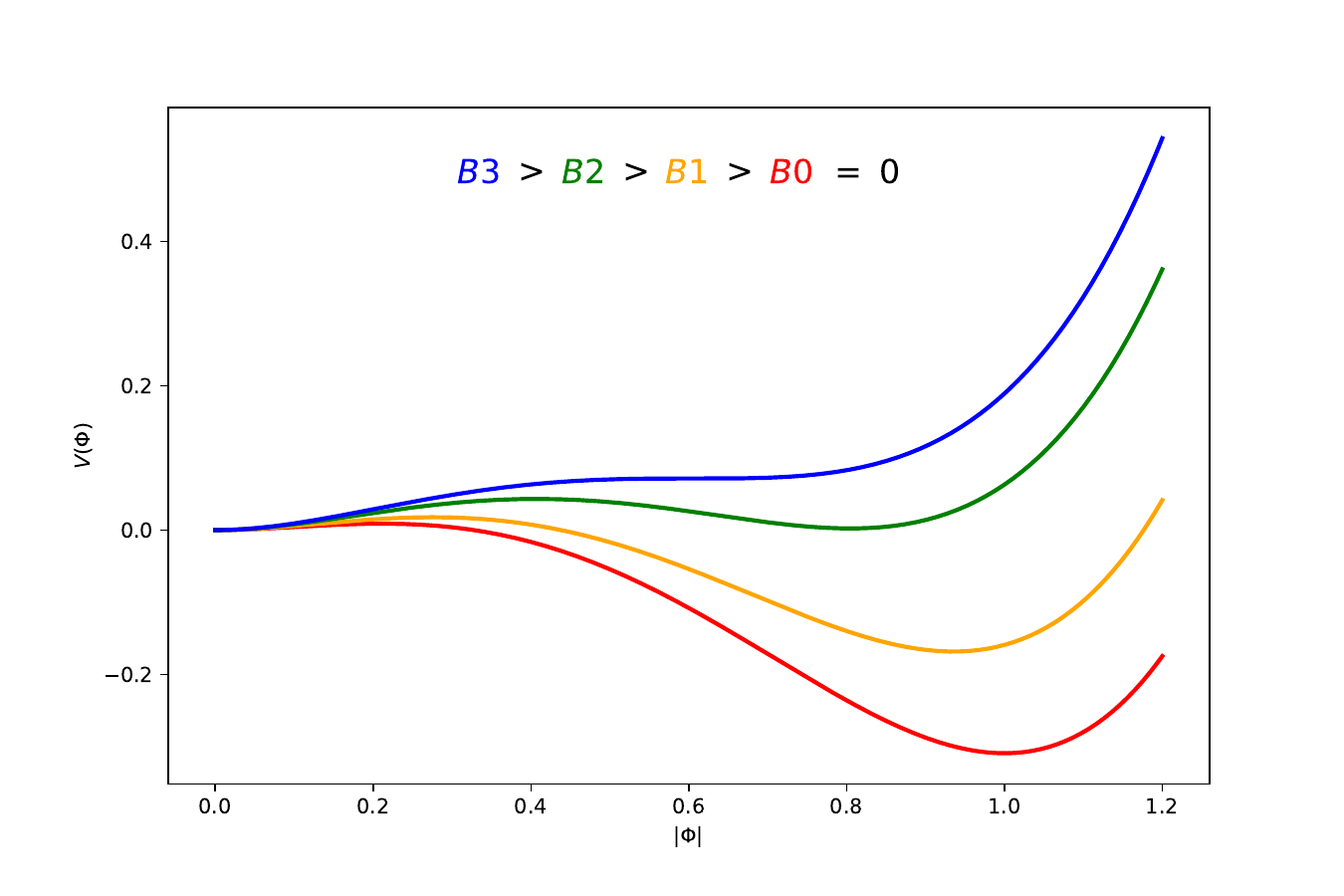}
    \caption{Schematic diagram of the additional contribution to the Higgs potential function \eqref{V2} from the term proportional to $(B_Y^\mathrm{ex})^2\Phi^\dagger\Phi$ obtained from covariant derivatives \eqref{D}. }
    \label{fig:VB}
\end{figure}

The potential $\mathcal{V} (\Phi)$ is expressed as
\begin{align}
    \mathcal{V} (\Phi) &= -\mu^2  \Phi^\dagger\Phi + A(\Phi^\dagger\Phi)^{3/2} + \lambda (\Phi^\dagger\Phi)^2\;.\label{V2}
\end{align}
It contains a potential barrier, as indicated by the red line in Fig. \ref{fig:VB}, that arises from particle physics beyond the SM.
Specially, during the PT, we have an extra contribution to the potential from the time-independent external MF through,
\begin{align}
\Delta V \supset \frac{g'^4}{4}Y_\mu^{\mathrm{ex}} Y^\mu_{\mathrm{ex}}\Phi^\dagger\Phi\;.
\end{align}
Before the electroweak PT, the electromagnetic field $\boldsymbol{B}^\mathrm{ex} = \nabla \times \boldsymbol{A}^\mathrm{ex}$ had not been well defined and was replaced by a hyperMF $\boldsymbol{B}^\mathrm{ex}_Y = \nabla \times \boldsymbol{Y}^\mathrm{ex}$. The relationship between them is $g'\boldsymbol{B}_Y^\mathrm{ex} = e\boldsymbol{B}^\mathrm{ex}$, here
$e=g\sin\theta_{\rm W}$ is the electric charge with $\theta_W$ being the Weinberg angle (Weak mixing angle). The term would cancel the quartic term in Eq.~\eqref{V2} to some extent and change the PT process accordingly for different MF distributions at different space-times.
For illustration, the influence of the external MF strength on the Higgs potential is demonstrated in Fig. \ref{fig:VB}. As the external MF increases, the value of $|\Phi|$ at the true vacuum continuously decreases, and the potential energy increases. If the external MF continues to increase, the true and false vacua degenerate, and ultimately, only one vacuum with $|\Phi|=0$ remains, with the electroweak symmetry being restored.

\begin{figure}[!htp]
    \centering
    \includegraphics[width=0.5\linewidth]{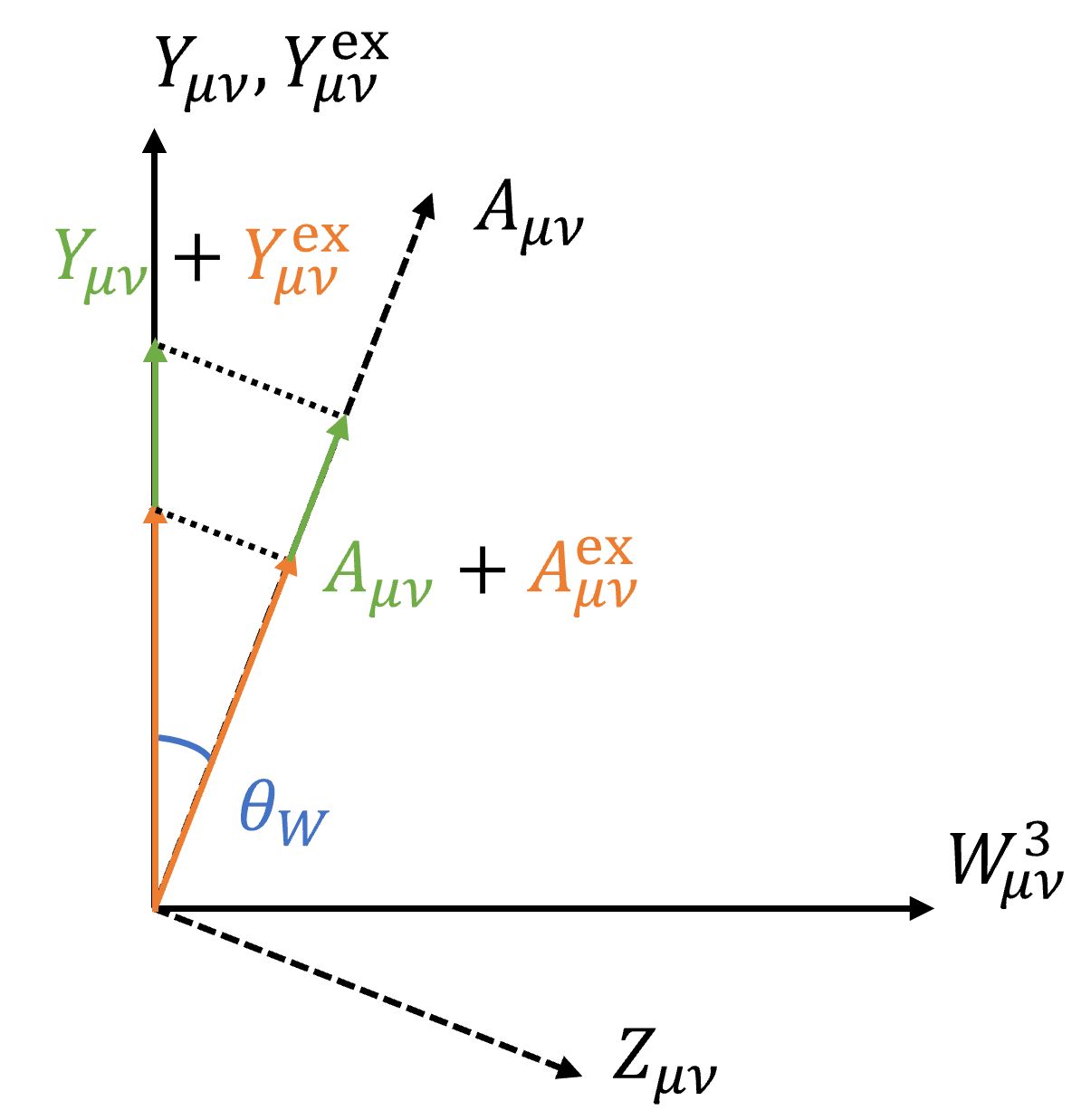}
    \caption{A graphic representation of the conversion from hyperMF $Y_{\mu\nu}$ into electromagnetic field $A_{\mu\nu}$ during the electroweak first order PT.}
    \label{fig:EWem}
\end{figure}

After the breaking of electroweak symmetry, the relationship between the electromagnetic field and the original gauge field is \cite{Zhang_Ferrer_Vachaspati_2017}
\begin{align}\label{eq:Amunu}
    A_{\mu\nu} =&\ W^3_{\mu\nu}\sin\theta_W + (Y_{\mu\nu}+Y_{\mu\nu}^
    \mathrm{ex})\cos\theta_W \\
    &\ - \ii \frac{2}{gv^2}[(D_\mu\Phi)^\dagger (D_\nu\Phi) - (D_\nu\Phi)^\dagger (D_\mu\Phi)]\sin\theta_W\;\notag\\
    =&\ W^3_{\mu\nu}\sin\theta_W + (Y_{\mu\nu}+Y_{\mu\nu}^
    \mathrm{ex})\cos\theta_W - S_{\mu\nu}\sin\theta_W \;,\notag
\end{align}
where the last term is the contribution of scalar current, which we denote as $S_{\mu\nu}$. Fig. \ref{fig:EWem} shows the transition of a hyperMF to an electromagnetic field during the first order PT. First-order PTs can generate MFs, serving as the seeds of the present-day cosmic MF~\cite{Di_2021}. When the external MF is stronger, it enters the scalar flow $S_{\mu\nu}$ (see Eq.~\eqref{eq:Amunu}) through covariant derivatives (see Eq.~\eqref{D}), further enhancing the MF generated by the PT. 

%

The chiral anomaly in electroweak theory gives the relationship between fermions and gauge fields \cite{Adler:1969gk, Bell:1969ts, tHooft:1976rip, tHooft:1976snw}:
\begin{align}
    \partial_\mu j_\mathrm{B}^\mu = N_\mathrm{g}\left[\frac{g^2}{16\pi^2}\mathrm{Tr}\,(W_{\mu\nu} \tilde W^{\mu\nu}) - \frac{g'^2}{32\pi^2}Y_{\mu\nu} \tilde Y^{\mu\nu}\right], 
\end{align}
where $N_\mathrm{g} = 3$ is the number of fermion generations, and the Hodge dual is $\tilde X^{\mu\nu}=\frac12\varepsilon^{\mu\nu\rho\sigma} X_{\rho\sigma}$. Integrating the above equation yields the change in baryon density
\begin{align}\label{eq:nbs}
    n_\mathrm{B} = N_\mathrm{g}\frac{\Delta N_\mathrm{CS}(t)}{V}
    = N_\mathrm{g}\frac{N_\mathrm{CS}(t)-N_\mathrm{CS}(0)}{V},
\end{align}
where $N_\mathrm{CS}(t)$ is the Chern-Simons number~\cite{Moore_2002}:
\begin{align}
    \frac{N_\mathrm{CS}(t)}{V} 
    &=\frac1V \frac{1}{32\pi^2}\varepsilon^{ijk}\int \mathrm{d}^3\boldsymbol{x} \biggl [ 
        -g'^2 (Y_i+Y^\mathrm{ex}_i) (Y_{jk}+Y^\mathrm{ex}_{jk})\nonumber\\
 &\ \ \ \ +g^2 \left ( W^a_i W^a_{jk}  -\frac{g}{3} \varepsilon^{abc} W^a_i W^b_j W^c_k \right )
   \biggr ]\;,\label{eq:ncs}
\end{align}
and $V=L^3$ is the lattice volume. 
The U(1) component of $N_\mathrm{CS}$ corresponds to the hypermagnetic helicity: 
\begin{align}
    h_{Y} = \frac{H_{Y}}{V} &= \frac1V \varepsilon^{ijk}\int \mathrm{d}^3\boldsymbol{x}\,
    (Y_i+Y_i^\mathrm{ex}) (Y_{jk}+Y_{jk}^\mathrm{ex}) \notag\\ 
    &= \frac1V \int \ \mathrm{d}^3\boldsymbol{x}\,
    (\boldsymbol{Y}+\boldsymbol{Y}^\mathrm{ex})
    \cdot (\boldsymbol{B}_Y+\boldsymbol{B}^\mathrm{ex}_Y),
\end{align}
up to a negative proportionality factor.
During first-order PT, bubble collisions trigger abrupt changes in the Higgs field $\Phi$ distributions in different space-time. This seeds the variation of the $g'\mathrm{Im}[\Phi^\dagger D_i\Phi]$ term in the $Y_i$ equation of motion, see Eqs.~(\ref{eom},\ref{gauss2}) and also Appendix~\ref{App:EOM}, driving rapid $Y_i$ field evolution that generates non-zero contributions in the Chern-Simons number and also the helicity from $\boldsymbol{Y}\cdot\boldsymbol{B}_Y$, $\boldsymbol{Y}\cdot\boldsymbol{B}^\mathrm{ex}_Y$ and  $\boldsymbol{Y}^\mathrm{ex}\cdot\boldsymbol{B}_Y$. Meanwhile, during the PT, the U(1)$_\mathrm{Y}$ field strength $Y_{\mu\nu} = \partial_\mu Y_\nu - \partial_\nu Y_\mu$ additionally contributes to electromagnetism through $A_{\mu\nu} \supset  Y_{\mu\nu} \cos\theta_W$~\cite{Di_2021}.


Bubble dynamics during the PT triggers abrupt changes in the Higgs field $\Phi$ distributions in different space-times, governed by the first equation in the following Eq.~\ref{eom}. The distributions of the nonvanishing Higgs field value when the PT starts yield the Higgs winding number $N_\mathrm{H}$ \cite{TUROK1991471}: 
\begin{align}
    N_\mathrm{H} = -\frac{1}{24\pi^2}\varepsilon^{ijk}\int\,\mathrm{d}^3x\Tr(\partial_i \varPhi^\dagger\partial_j \varPhi\partial_k \varPhi^\dagger \varPhi),
\end{align}
with \cite{Mou_Saffin_Tranberg_2018}
\begin{align}
    \varPhi = \frac{1}{|\Phi|}(\ii \sigma^2\Phi^*,\Phi) 
\end{align}
being a unitary matrix, $N_\mathrm{H}$ would grow along with the changing of the Chern-Simons number and stabilize when the PT is complete, see the following section. 

During the PT, the baryon (and lepton) asymmetry of the Universe can be calculated through the relationship between the change in baryon (and lepton) number and the variation of the Chern-Simons number:
\begin{align}
    \eta_\mathrm{B}(t) = \frac{n_\mathrm{B}}{s}=\eta_\mathrm{L}(t) = \frac{n_\mathrm{L}}{s}=\frac{45N_\mathrm{g}}{2\pi^2 g_{*S}T^3}\frac{\Delta N_\mathrm{CS}(t)}{V} \;.\label{eq:etaB}
\end{align}
Here, $g_{*S}=106.75$ is the effective number of degrees of freedom in entropy before the PT. 
The time-varying hypermagnetic helicity in the early Universe enters the anomalous process, leading to baryon and lepton number asymmetry while preserving $\mathrm{B}-\mathrm{L}$ conservation~\cite{Anber:2015yca, Giovannini:1999wv, Giovannini:1999by, Boyarsky:2011uy, Fujita:2016igl, Brustein:1998du, Boyarsky:2011uy, PhysRevLett.108.031301, Joyce:1997uy}:

\begin{equation}
    \begin{aligned}
        &\frac{\mathrm{d} \eta^Y_B}{\mathrm{d} t}=\frac{\mathrm{d} \eta^Y_L}{\mathrm{d} t}=-\frac{45N_\mathrm{g}}{2\pi^2g_{*S}T^3}\left(\frac{g'^2}{16\pi^2}\frac{\mathrm{d}h_Y}{\mathrm{d}t}\right)\;.
    \end{aligned}\label{etabY}
\end{equation}
Temporal evolution of the Chern-Simons number and hypermagnetic helicity thus originates from interactions between the preexisting external helical hypermagnetic field and PT-generated magnetic fields. The net baryon number generation is subsequently quenched after the PT completes.


\section{The Simulation Setup}\label{App:setup}
Under the temporal gauge $W_0=Y_0=Y^\mathrm{ex}_0 = 0$, the equations of motion are: 
\begin{align}
	\partial_0^2 \Phi\hphantom{{}_i} =& D_i D_i \Phi - 
	\frac{\partial \mathcal{V}}{\partial \Phi^\dagger}  \;,\notag\\ 
	\partial_0^2 W^a_i =&-\partial_k W^a_{ik}  \notag 
    - g\epsilon^{abc}W^b_k W^c_{ik} + g\text{Im}[\Phi^\dagger \sigma^a(D_i\Phi)] 
	\notag\;, \\  
\partial_0^2 Y_i =& -\partial_k Y_{ik}+g'\text{Im}[\Phi^\dagger(D_i\Phi)]\;,
\label{eom}
\end{align}
along with two Gauss constraints,
\begin{align}
	&\partial_0 \partial_j Y_j -g'\text{Im}\big[ \Phi^\dagger \partial_0 \Phi \big] = 0\;, \notag \\ 
	& \partial_0 \partial_j W^a_j + g \epsilon^{abc} W^b_j \partial_0 W^c_j 
        -g\text{Im} \big[ \Phi^\dagger \sigma^a \partial_0 \Phi \big] =0.
\label{gauss2}
\end{align}
We here employ natural units where $\hbar=c=k_B=1$. We neglect the expansion of the universe and accompanied temperature variation, and also the evolution of background MFs during the first-order PT, since it is a rapid process. Please refer to \textit{Equations of Motion on Lattice} of Appendix.~\ref{App:EOM} for the discretized equations of motion and Gaussian constraints on the lattice.

We use the thermal fluctuation spectrum under thermal equilibrium to describe the Higgs field $\Phi$ and its conjugate momentum field $\Pi$ before PT: 
\begin{align}
    \mathcal{P}_{\Phi_i}(k) = \frac{1}{\omega_k} \frac{1}{e^{{\omega_k}/T} - 1}, \qquad
    \mathcal{P}_{\Pi_i}(k) = \frac{\omega_k}{e^{{\omega_k}/T} - 1},
\end{align}
where $\omega_k = \sqrt{k^2 + m^2_\mathrm{eff}}$. $k$ and $m_\mathrm{eff}$ are the physical momentum and effective mass of the Higgs field.

In the continuum, we have
\begin{align}
    \langle\Phi_i(\boldsymbol{k})\Phi_j(\boldsymbol{k}')\rangle &= (2\pi)^3\mathcal{P}_{\Phi_i}(k)\delta(\boldsymbol{k}-\boldsymbol{k}')\delta_{ij}, \label{eq:phithermal}\\
    \langle\Pi_i(\boldsymbol{k})\Pi_j(\boldsymbol{k}')\rangle &= (2\pi)^3\mathcal{P}_{\Pi_i}(k)\delta(\boldsymbol{k}-\boldsymbol{k}')\delta_{ij}, \\
    \langle\Phi_i(\boldsymbol{k})\Pi_j(\boldsymbol{k}')\rangle &= 0.
\end{align}



Converting it into a discrete form on the lattice, we get
\begin{align}
    \langle|\Phi_i(\boldsymbol{k})|^2\rangle = \left(\frac{N}{\Delta x}\right)^3\mathcal{P}_{\Phi_i}(k), \qquad
     \langle\Phi_i(\boldsymbol{k})\rangle = 0, \\
     \langle|\Pi_i(\boldsymbol{k})|^2\rangle = \left(\frac{N}{ \Delta x}\right)^3\mathcal{P}_{\Pi_i}(k), \qquad
     \langle\Pi_i(\boldsymbol{k})\rangle = 0,
\end{align}
where $N^3$ denotes the total number of lattice points and $\Delta x$ is the physical lattice spacing. $\Phi_i(\boldsymbol{k})$ and $\Pi_i(\boldsymbol{k})$ satisfy the point independent Gaussian distribution in the momentum space, which contains all modes from infrared truncation $k_\mathrm{IR} = 2\pi/(N\Delta x)$ to the ultraviolet cutoff $k_\mathrm{CutOff}$. For the gauge fields, we set their initial values to $0$. However, to satisfy the Gausan constraint (Eq.~\eqref{gauss2}), we must assign an initial value to the conjugate momentum field of the gauge field according to certain rules. For specific details on {\it Initialization of gauge fields}, see Appendix~\ref {App:init}. 

When PT occurs, the Higgs field quantum tunnels from the false vacuum $|\Phi|\simeq0$ to the true vacuum $|\Phi|=v$, nucleating a bubble at $t=t_0$. In the simulation, each lattice site near the false vacuum at each time step has a probability $p_\mathrm{bubble}$ to nucleate a vacuum bubble, which has the profile
\begin{align}
    \Phi(t_0, r) &= \frac v2\left[1-\tanh\left(\frac{r-R_0}{l_\mathrm{w}}\right)\right]
    \begin{pmatrix}
        0 \\ 1
    \end{pmatrix} ,\label{bubbleprofile} \\
    \dot\Phi(t_0, r) &= 0,
\end{align}
where $r$ denotes the distance from bubble center, $R_0$ denotes the radius of bubble, and $l_\mathrm{w}$ denotes the bubble wall thickness. Then, the nucleated bubbles expand and collide with each other to propel the PT process. 

Considering the correlation length of the primordial MFs, generated through inflation~\cite{Turner:1987bw,Adshead:2016iae,Martin:2007ue,Kanno:2009ei,Ratra:1991bn}, the first-order PT~\cite{PhysRevLett.51.1488, Di_2021, Enqvist:1993np,Yang_2022,Vachaspati:1991nm,Baym:1995fk,Grasso:1997nx,Quashnock:1988vs}, we consider two types of external hyperMFs with distinct correlation lengths, defined as 
$\lambda_{BY} = \int \mathrm{d}k\, k^{-1}E_{BY}(k)\ \Big/ \int \mathrm{d}k\, E_{BY}(k)$, 
where $\int \mathrm{d}k\, E_{BY}(k) = \int \mathrm{d}^3\boldsymbol{x}\, \frac12 [\boldsymbol{B}^\mathrm{ex}_Y(\boldsymbol{x})]^2/V$. 
The infinite correlation length hyperMF with helicity is set by $Y^{\mathrm{ex}\mu} = (0,\ 0,\ xB_Y^\mathrm{ex},\ h_\mathrm{factor}LB_Y^\mathrm{ex})$, where $h_\mathrm{factor}\in[-1,1]$, and $L = N\Delta x$ represents the edge length of the lattice box. 
For hyperMFs with finite correlation lengths, we adopt the following form in Fourier space~\cite{Brandenburg_2020}:
$   \Tilde{B}_{Yi}^\mathrm{ex}(\boldsymbol{k}) = B_\mathrm{ini}\Theta(k-k_\mathrm{UV})\left(\delta_{ij}-\hat{k}_i\hat{k}_j-\ii\sigma_\mathrm{M}\varepsilon_{ijl}\hat{k}_l\right)g_j(\boldsymbol{k})k^n$. 
Here $\Theta(k-k_\mathrm{UV})$ is the Heaviside function, ensuring the hyperMF has a finite correlation length comparable to the mean bubble separation $R_*=(V/N_\mathrm{bubble})^{1/3}$ during the percolation process of vacuum bubbles in the PT. $\boldsymbol{g}(\boldsymbol{k})$ is the Fourier transform of a Gaussian-distributed random vector field that is $\delta$-correlated in all three dimensions. The helicity degree is controlled by $\sigma_\mathrm{M}=0,1,-1$, corresponding to non-helical, maximally positive, and maximally negative helicities, respectively. 
The specific implementation of hyperMFs on the lattice is detailed in the {\it Hypermagnetic Field on Lattice} section of the Appendix.~\ref{sec:hyMFlat}. 

We conduct numerical simulations on two different sizes of three-dimensional lattices, with lattice spacing of $\Delta x$ and periodic boundary conditions:
\begin{itemize}
    \item $N^3 = 128^3$: We simulate a homogeneous MF and a spectral distribution MF with $\lambda_B\sim R_*/4\sim R_0$, where $R_0$ is the initial vacuum bubble radius. We conduct 20 simulations with $n_\mathrm{timestep}=10000$ for each MF strength and helicity setting. 
    \item $N^3 = 512^3$:  We simulate the spectral distribution MF with $\lambda_B\sim R_*$. Due to computational constraints, we limit the simulations to 4 runs with $n_\mathrm{timestep}=3600$ for each combination of $\sigma=\pm 1$ and hyperMF strength.
\end{itemize}
The model parameters and lattice setup in the simulation are presented in Tab. \ref{tab:parameter}.
\begin{table}[!htp]
    \centering
    \begin{tabular}{c|c|c|c|c|c|c|c|c}
        \hline\hline
        $v$ &$\Delta x$& $\Delta t$&$\mu^2$ & $A$ &$\lambda$ & $R_0$ & $l_\mathrm{w}$ & $p_\mathrm{bubble}$ \\
        \hline
        $1$ &0.2&0.04& 0.68 & 2.59 & 1.60 &$13\Delta x$ &$6\Delta x$&$5\times10^{-8}$ \\
        \hline\hline
    \end{tabular}
    \caption{Model parameters and setup in the simulation. }
    \label{tab:parameter}
\end{table}




\section{Numerical results}

In this section, we numerically study the following ingredients in the background of the hyperMF: the PT speed, the phenomena of the Ambj\o rn-Oleson condensation, the electroweak sphaleron rate, and the baryon and lepton number generation through the chiral anomaly. 

\subsection{Phase Transition speed} \label{app:PT}

\begin{figure}[!htp]
    \centering
    \includegraphics[width=0.4\textwidth]{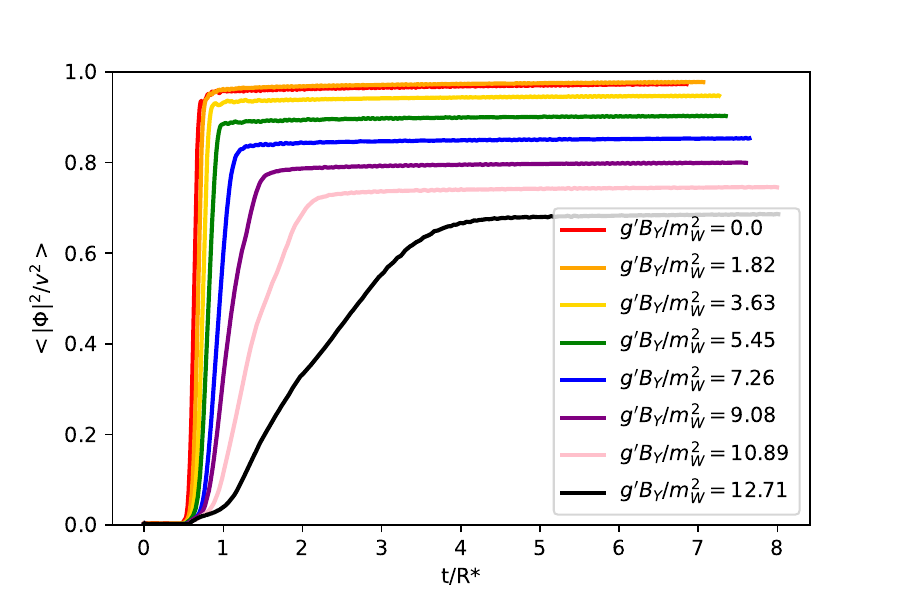}
    \includegraphics[width=0.4\textwidth]{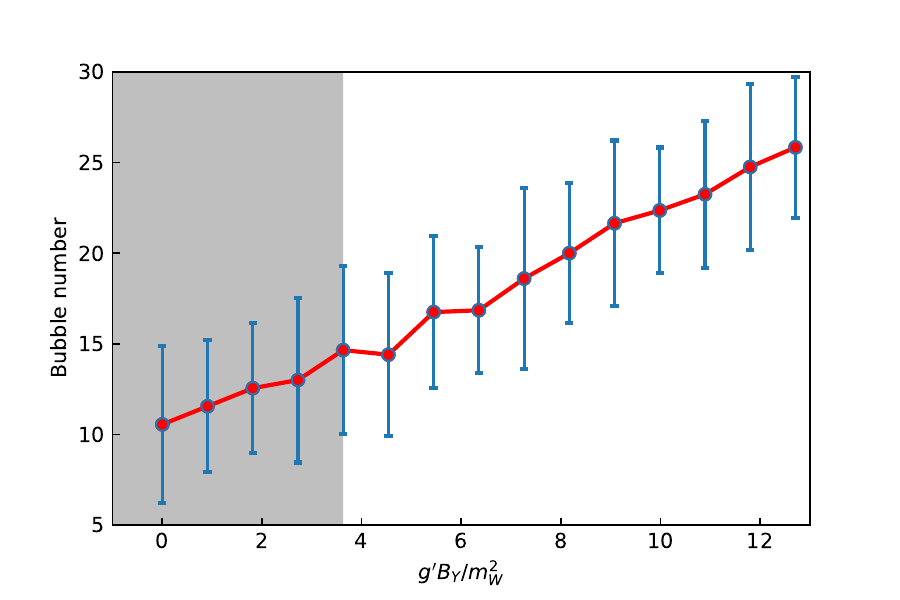}
    \caption{Top: Evolution of $\Phi^2$ over time under non-helical homogeneous hyperMF. Bottom: The number of bubbles nucleated under non-helical homogeneous hyperMF. The gray area indicates the hyperMF strength where Higgs condensation has not yet occurred.}
    \label{fig:Phi2vsNb}
\end{figure}

\begin{figure*}[!htp]
    \centering
    \includegraphics[scale=0.4]{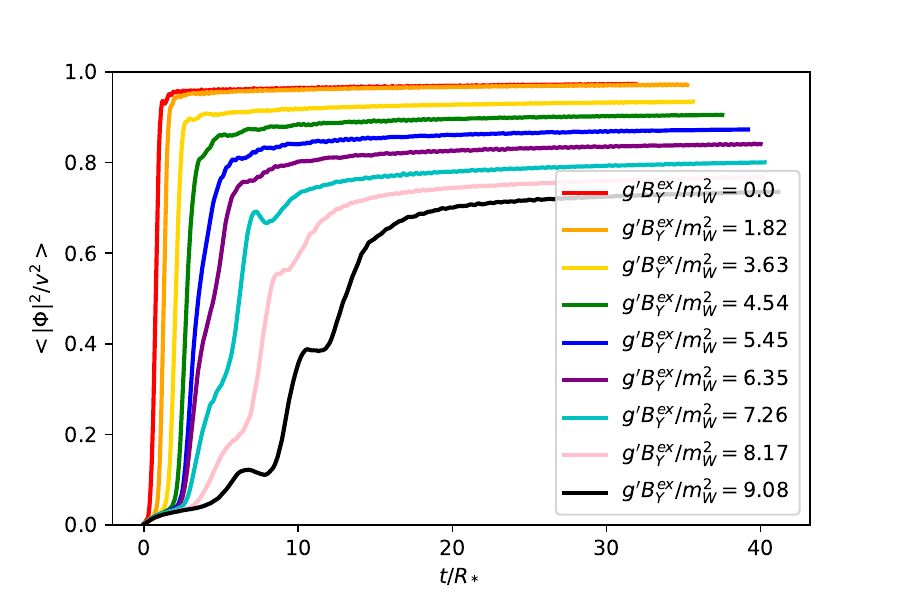}
    \includegraphics[scale=0.4]{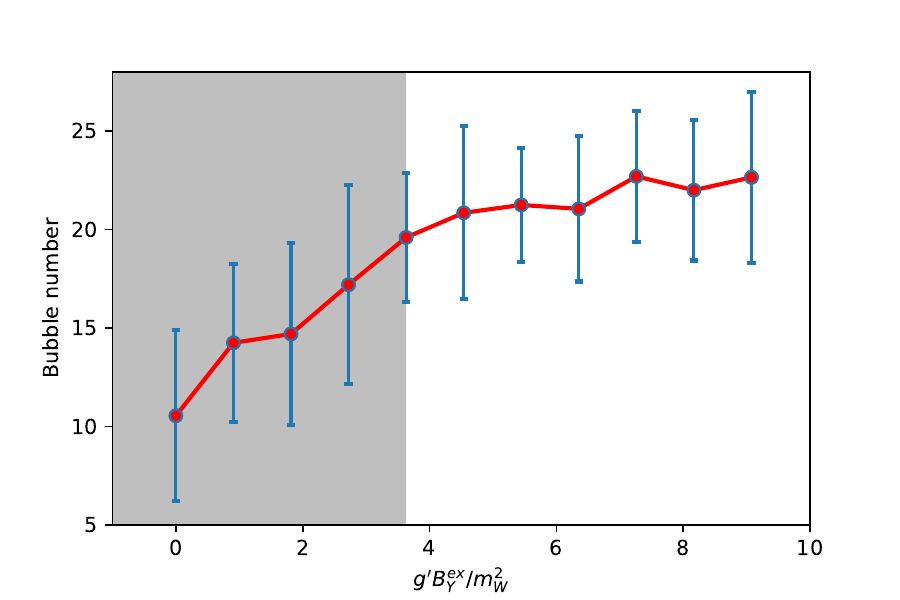}\\
    \includegraphics[scale=0.4]{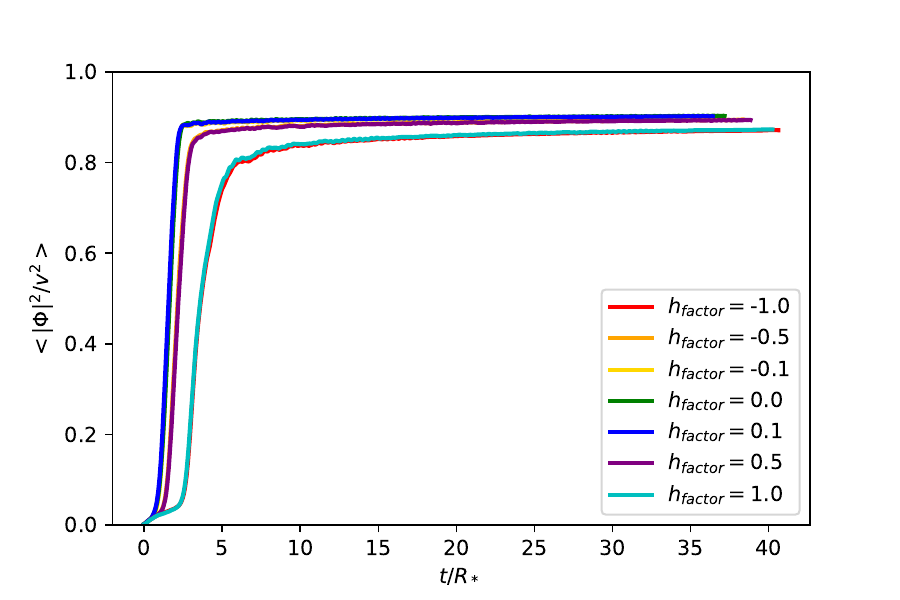}
    \includegraphics[scale=0.4]{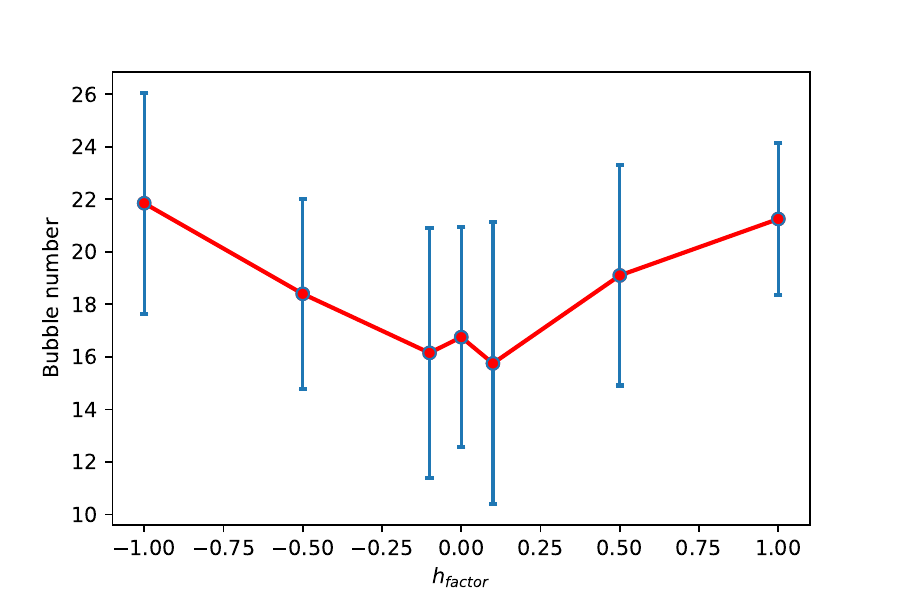}
    \caption{Left: Evolution of $\Phi^2$ over time under different helical homogeneous hyperMF strength with $h_\mathrm{factor}=1$ (Top) and under the same homogeneous hyperMF strength ($g'B_Y^\mathrm{ex}/m_W^2 = 5.45$) and different $h_\mathrm{factor}$ (Bottom). Right: The number of bubbles nucleated under different MF strengths with $h_\mathrm{factor}=1$ (Top) and under different $h_\mathrm{factor}$ with homogeneous hyperMF strength being fixed at $g'B_Y^\mathrm{ex}/m_W^2 = 5.45$ (Bottom). The gray area indicates the magnetic field strength where Higgs condensation has not yet occurred.}
    \label{fig:phi2vsNBhel}
\end{figure*}

\begin{figure*}[!htp]
    \centering
    \includegraphics[scale=0.4]{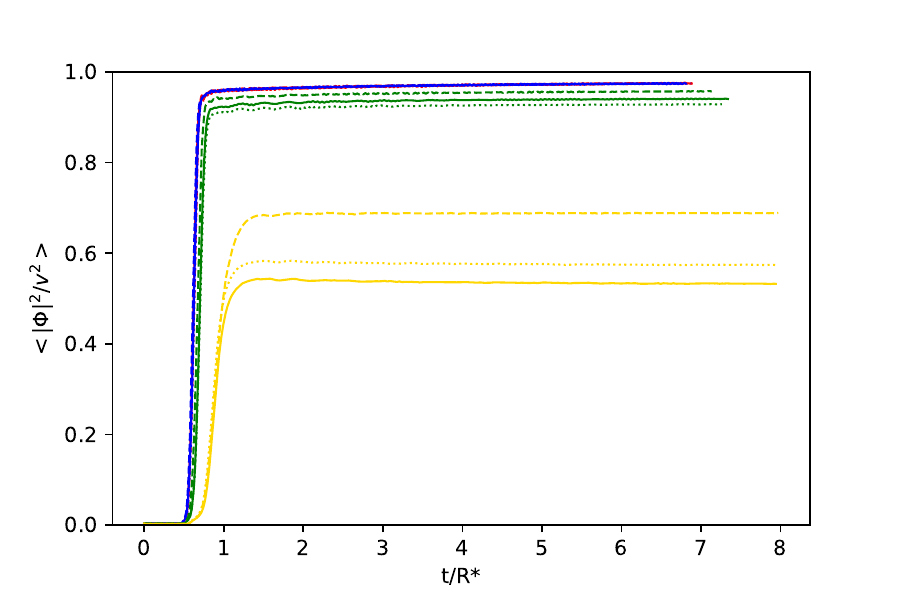}
    \includegraphics[scale=0.4]{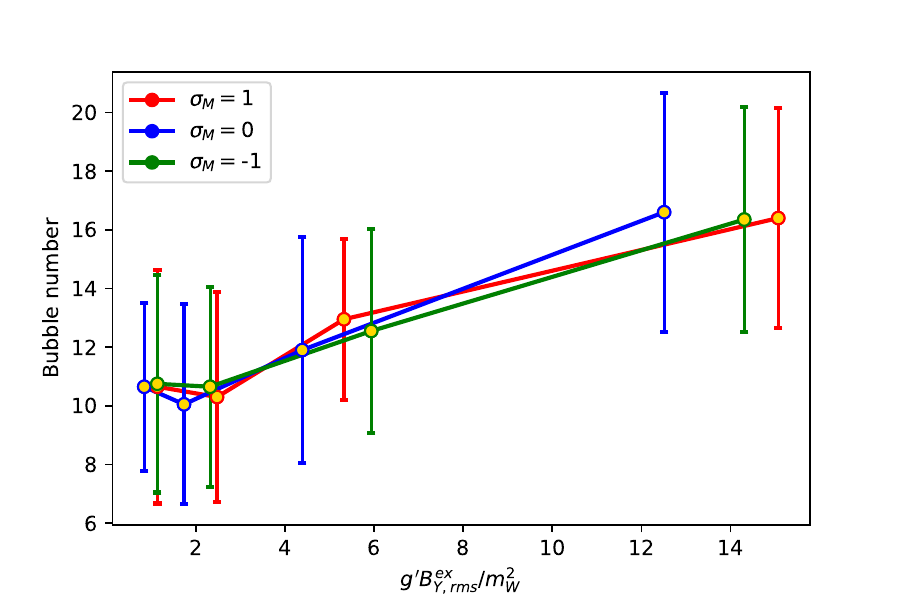}\\
    \includegraphics[scale=0.4]{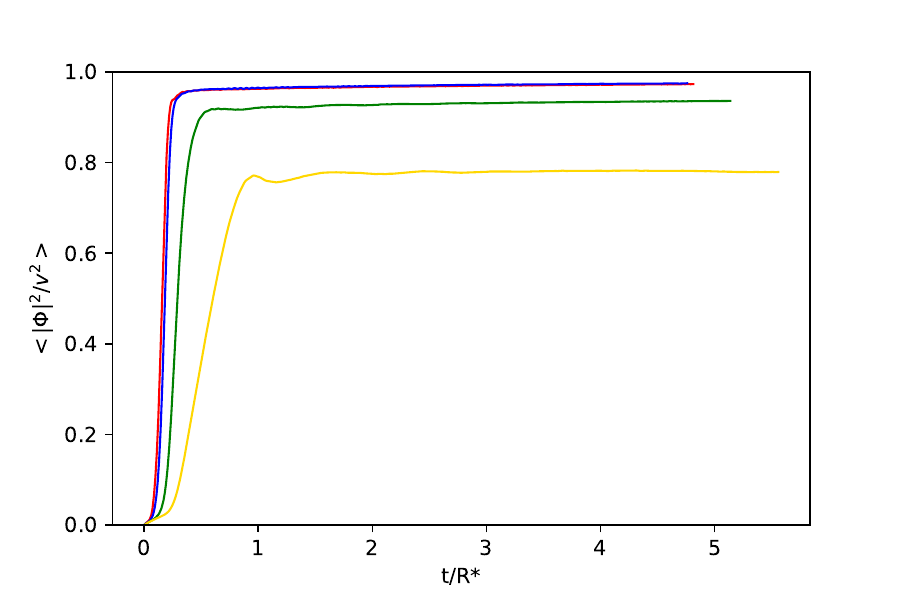}
    \includegraphics[scale=0.4]{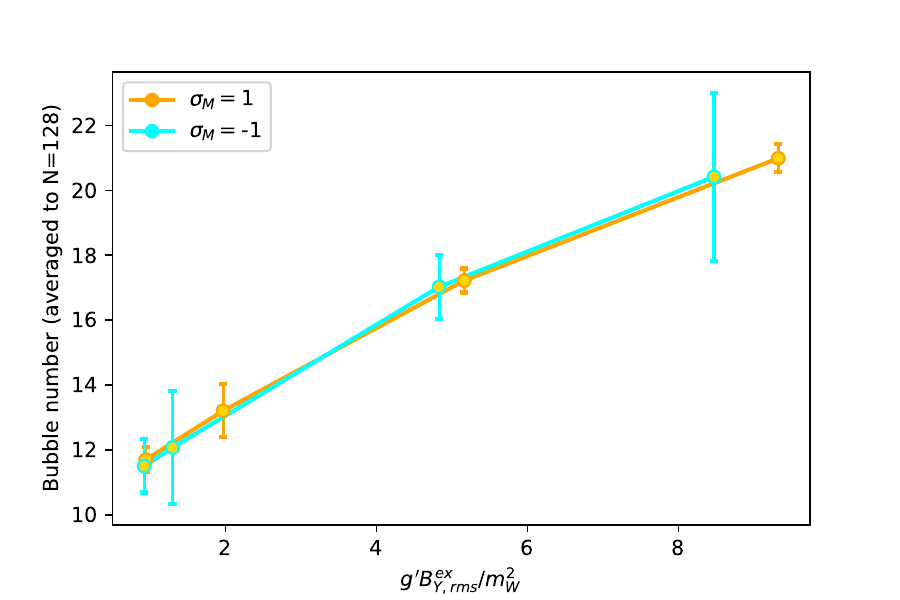}
    \caption{Left: Evolution of $\Phi^2$ over time under different spectrum index $n$ or different hyperMF with spectral distribution. $n = 0,1,2,3$ corresponds to yellow, green, blue, and red lines, respectively. The solid line indicates $\sigma_\mathrm{M}=1$, the dotted line indicates $\sigma_\mathrm{M}=-1$, and the dashed line indicates $\sigma_\mathrm{M}=0$, respectively. Right: The number of bubbles nucleated under different spectrum-distributed hyperMF strengths. The case of $N=128$ is shown in the top column, while the case of $N=512$ is shown in the bottom. Due to the spatial non-uniformity of the hyperMF in the spectral distribution, we use the root mean square hyperMF stength $B_{Y,\mathrm{rms}}^\mathrm{ex} = \sqrt{2\rho_{BY}^\mathrm{ex}}$ as the horizontal axis, where $\rho_{BY}^\mathrm{ex}$ is the volume averaged hyperMF energy density.}
    \label{fig:SBphibubble}
\end{figure*}

Fig. \ref{fig:Phi2vsNb} shows the change of the Higgs field over time and the number of bubbles under different homogeneous non-helical hyperMF strengths. It can be seen that when the MF strength increases, the PT speed becomes slower and more bubbles are nucleated, which reflects that the strong MF affects the electroweak vacuum. When a bubble is nucleated, the strong MF acts rapidly on the true vacuum area inside the bubble. This not only reduces the vacuum expectation value of the Higgs field but also slows down the expansion speed of the bubble, causing the false vacuum to occupy the space of the universe for a longer period, which further results in the creation of more vacuum bubbles.

The effects of homogeneous helical hyperMF on the PT speed and the number of bubbles are similar to those without helicity. As shown in the top panels of Fig. \ref{fig:phi2vsNBhel}, the helical homogeneous hyperMF makes the PT speed slower and also makes the final stable field value smaller. This shows that the speed of bubble expansion is slowed down in the hyperMF with helicity, and the bubble expansion speed is further reduced compared with the hyperMF without helicity. The helicity will also slightly affect the PT process. As shown in the bottom panels in Fig. \ref{fig:phi2vsNBhel}, as the initial helicity increases, the PT speed will decrease slightly, and the final value of $\langle|\Phi|^2\rangle$ will also become smaller. This is just as we stated before, the speed at which the bubble expands is further slowed down by the helicity. 


From Fig. \ref{fig:SBphibubble}, it can be seen that the correlation length has an impact on the PT speed. When the correlation length is small, the ``transition" region between regions with high MF strength and regions with low MF strength is small, and the PT will quickly complete in the regions with low MF strength, while in the regions with high MF strength, the PT will not occur at all. This leads to a PT speed in this situation that is not significantly different from the time difference without a MF, but $\langle|\Phi|^2\rangle$ will decrease. On the contrary, MFs with larger correlation lengths have larger ``transition" regions, where the MF strength is moderate and PT can occur here, but at a slower speed. The extreme case is a homogeneous MF with an infinite correlation length.

\subsection{Ambj\o rn-Oleson Condensation}

When the process of electroweak symmetry breaking occurs in a MF background, the configuration of the broken phase field will be closely related to the MF strength. 
There exist two critical MFs as follows:
\begin{align}
    g'B_{Yc1} = eB_{c1} = m_W^2\;,\label{eq:BYc1}\\
    g'B_{Yc2} = eB_{c2} = m_H^2\;.
\end{align}
As the external MF increases, the Higgs and gauge fields exhibit different behaviors \cite{Nielsen:1978rm, PhysRevD.45.3833, SALAM1975203, LINDE1976435,Ambjorn_1989, Ambjorn:1988fx, Ambjorn:1988gb, Ambjorn:1988tm, Ambjorn:1989bd}:
\begin{itemize}
    \item When $B<B_{c1}$, the MF has no effect on the broken phase. As symmetry breaking progresses, the Higgs field will gradually fall into the broken phase and stabilize there. 
    \item When $B_{c1}<B<B_{c2}$, the original electroweak vacuum becomes unstable due to the emergence of an imaginary part in the vacuum energy. To compensate for this instability, the SU(2)$_\mathrm{L}$ gauge field forms a hexagonal vortex line structure (with the direction of the line following the direction of the external MF). This phenomenon is known as the Ambj\o rn-Oleson condensation. Since the SU(2)$_\mathrm{L}$ gauge field couples with the Higgs field through covariant derivatives \eqref{D}, the Higgs field also produces a similar hexagonal vortex line structure, resulting in a decrease in the vacuum expectation value.
    \item When $B>B_{c2}$, the electroweak symmetry will be restored. This means that the vortex line structure of gauge fields and the Higgs field will disappear, and the vacuum expectation value of the Higgs field will return to 0. At this point, the external MF will be replaced by the U(1)$_\mathrm{Y}$ gauge field.
\end{itemize}

\begin{figure}[!htp]
    \centering
    \includegraphics[width=0.2\textwidth]{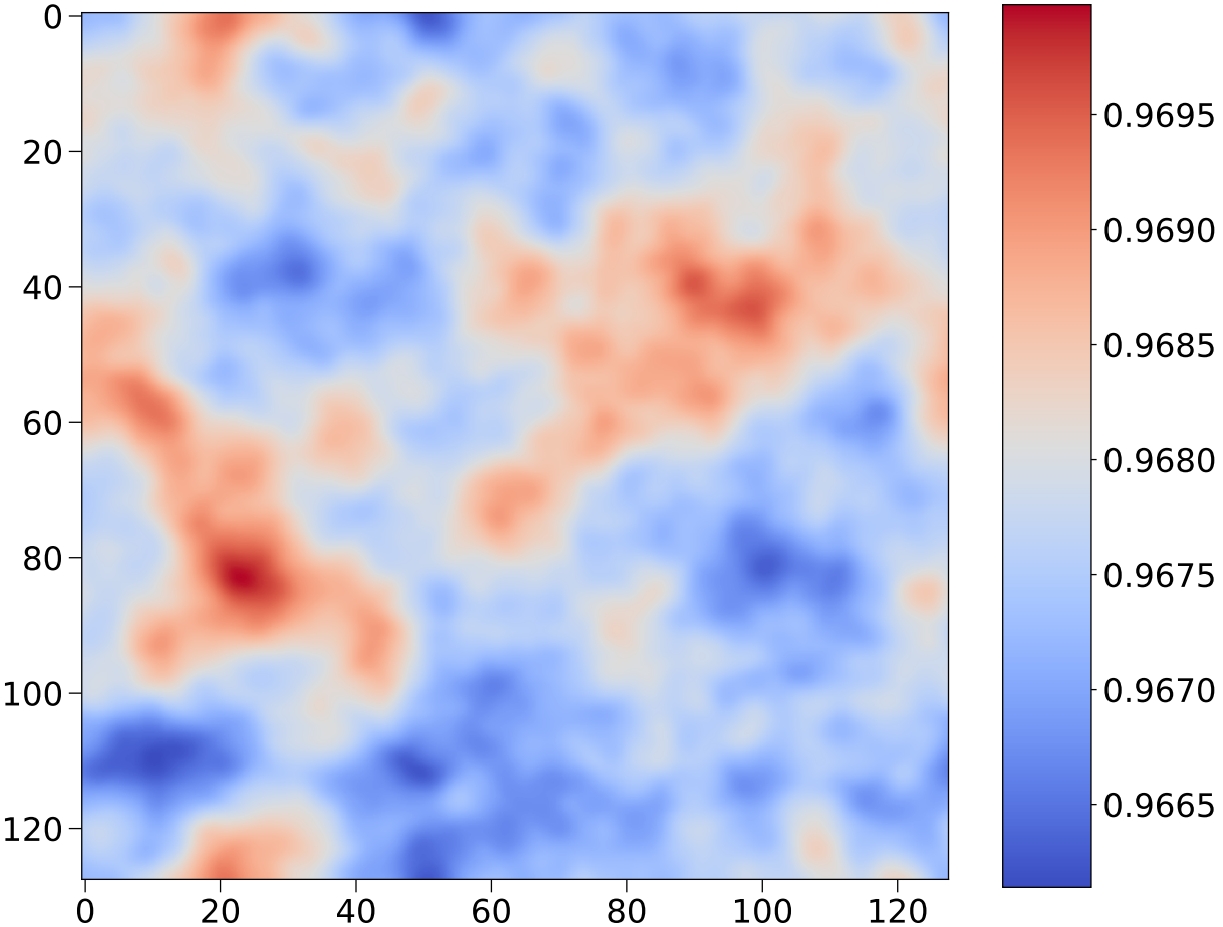}
    \includegraphics[width=0.2\textwidth]{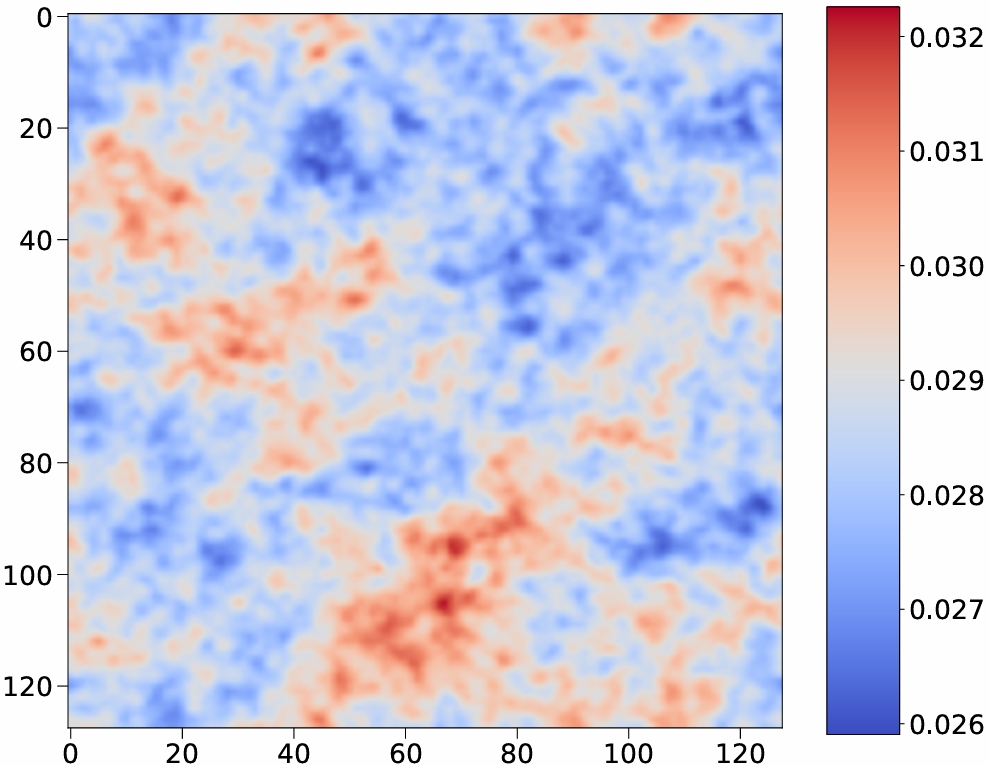}
    \includegraphics[width=0.2\textwidth]{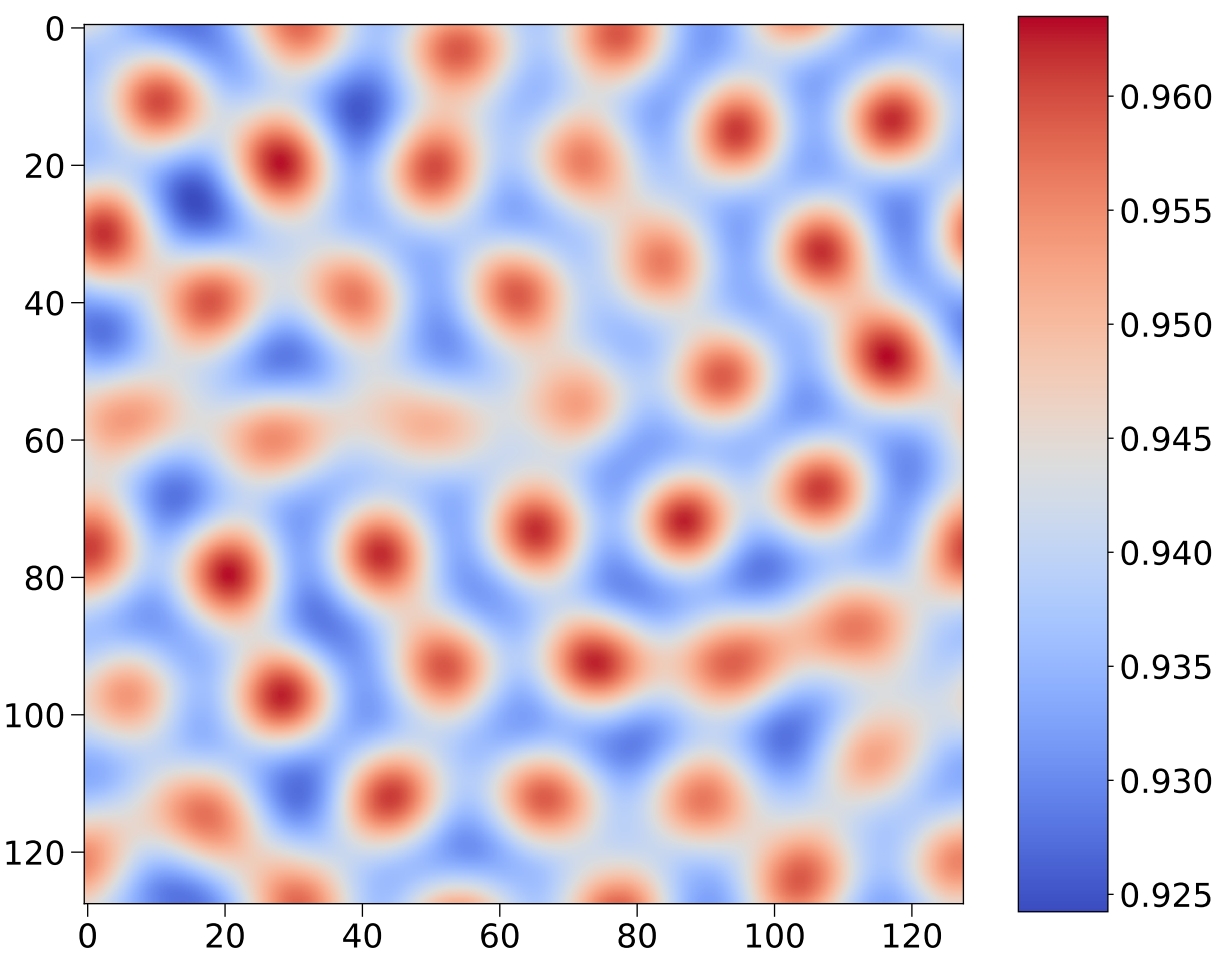}
    \includegraphics[width=0.2\textwidth]{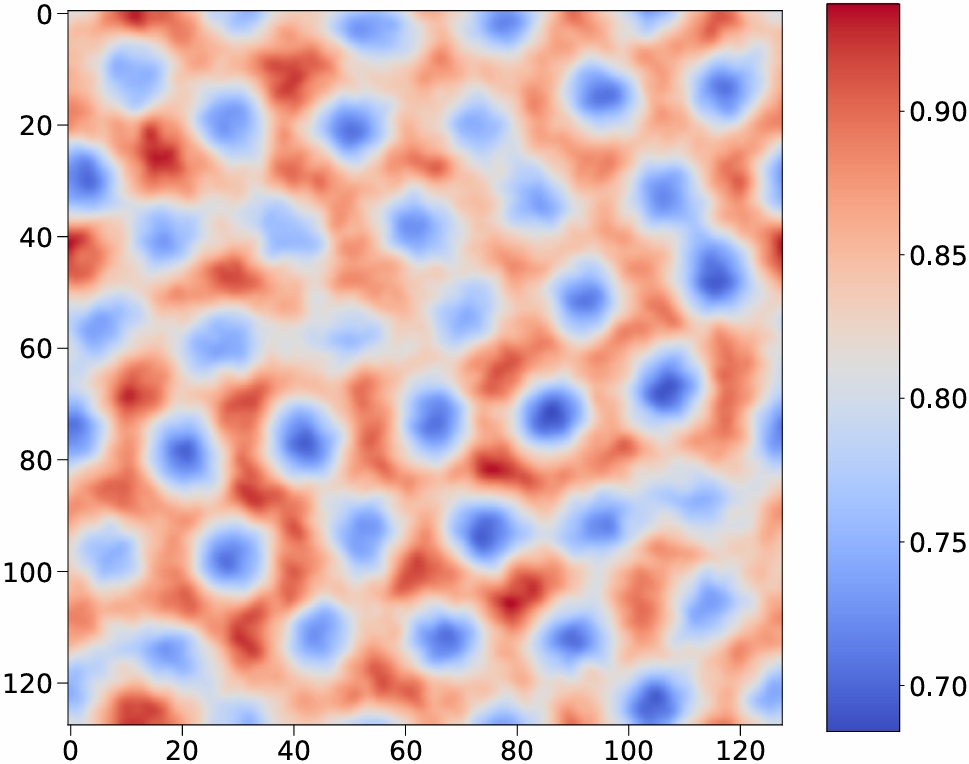}
    \includegraphics[width=0.2\textwidth]{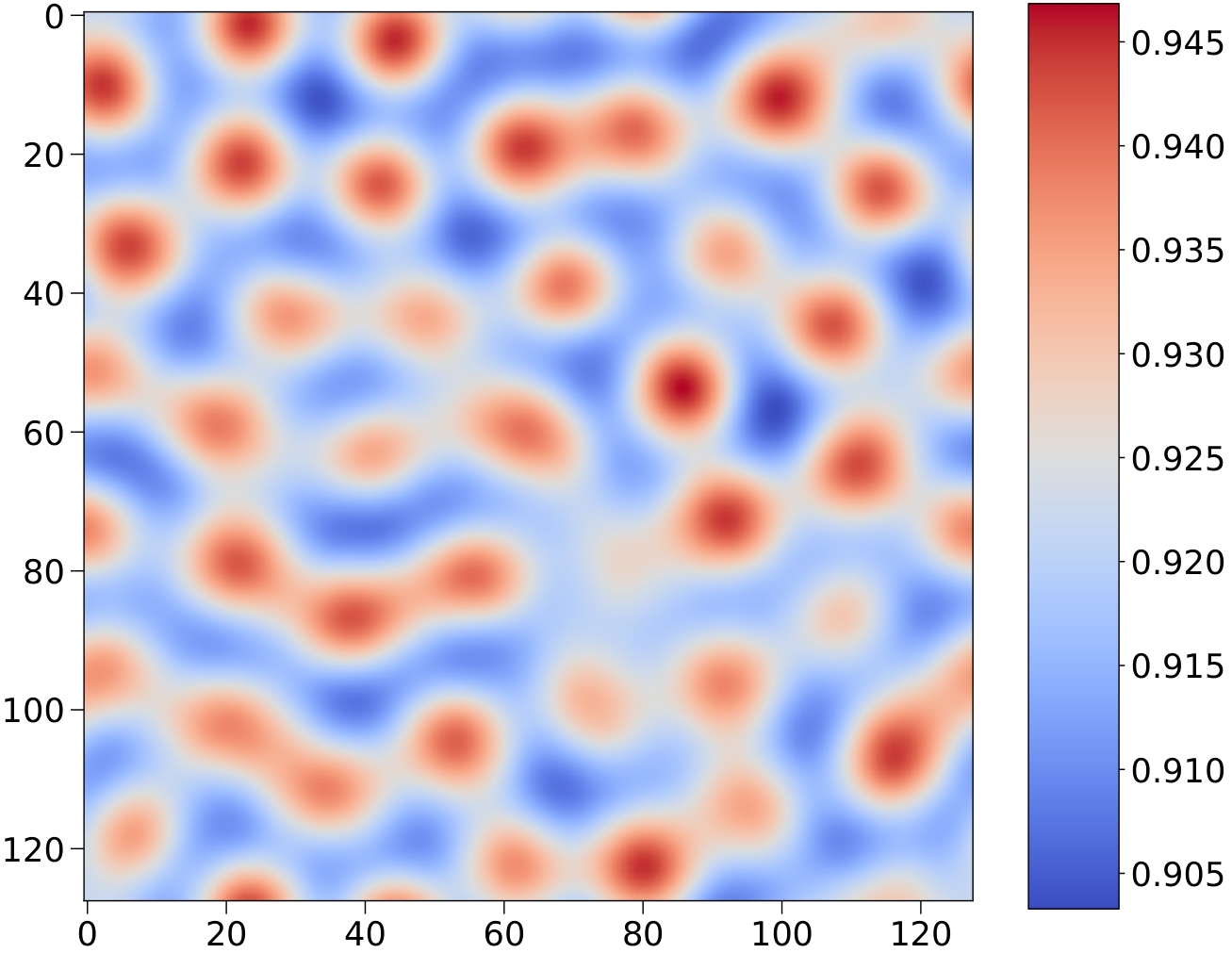}
    \includegraphics[width=0.2\textwidth]{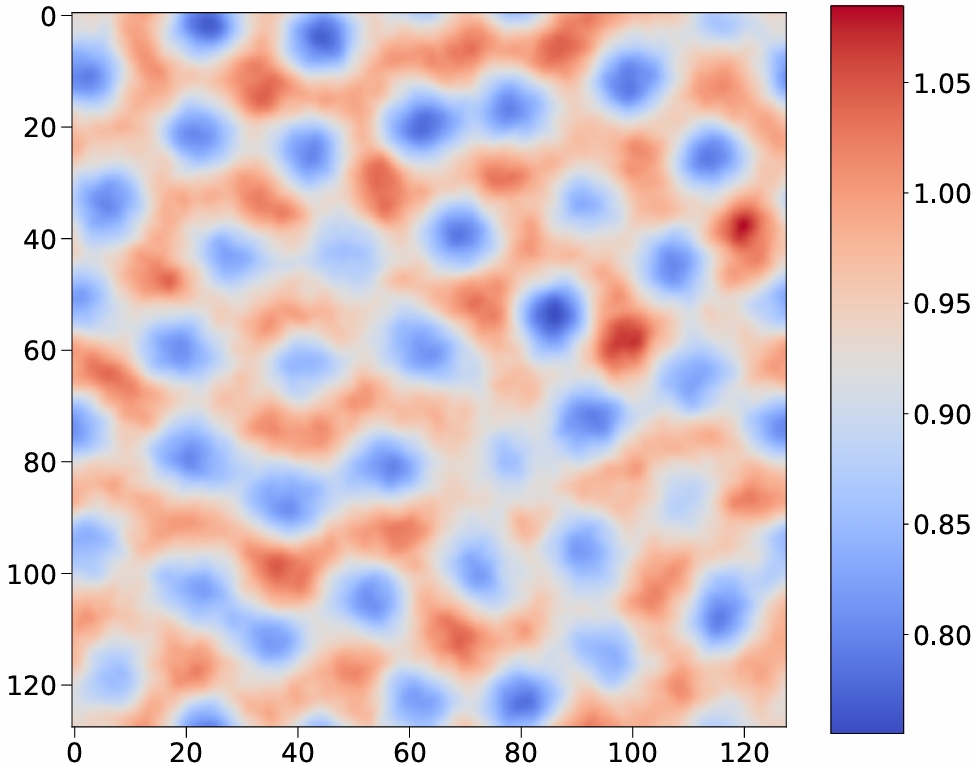} 
    \caption{$\overline {\Phi^2}$(Left) and $\overline {B^2}$(Right) under different homogeneous hyperMFs. The top two plots indicate the case where $g'B_Y^\mathrm{ex}/m_W^2=0$, and the middle and bottom two plots are the scenarios with non-helical and helical homogeneous hyperMFs ($h_{\rm factor}=1$) with $g'B_Y^\mathrm{ex}/m_W^2=3.63$ where the Higgs condensations appear.
 }
    \label{fig:Higgs2}
\end{figure}

In the case of cross-over, the Ambj\o rn-Oleson condensation phenomenon has been verified through lattice simulation methods \cite{Chernodub_2023}. 
Ref.~\cite{Kajantie:1998rz} studied the electroweak PT dynamics under external MF and didn't observe the Ambjørn-Olesen phase, which might be due to the quantum fluctuations there. 
We will demonstrate that Ambj\o rn-Oleson condensation indeed can occur in the context of first-order PT, and it bears little relation to the helicity of the MF.

\begin{figure*}[!htp]
    \centering
    \includegraphics[width=0.3\linewidth]{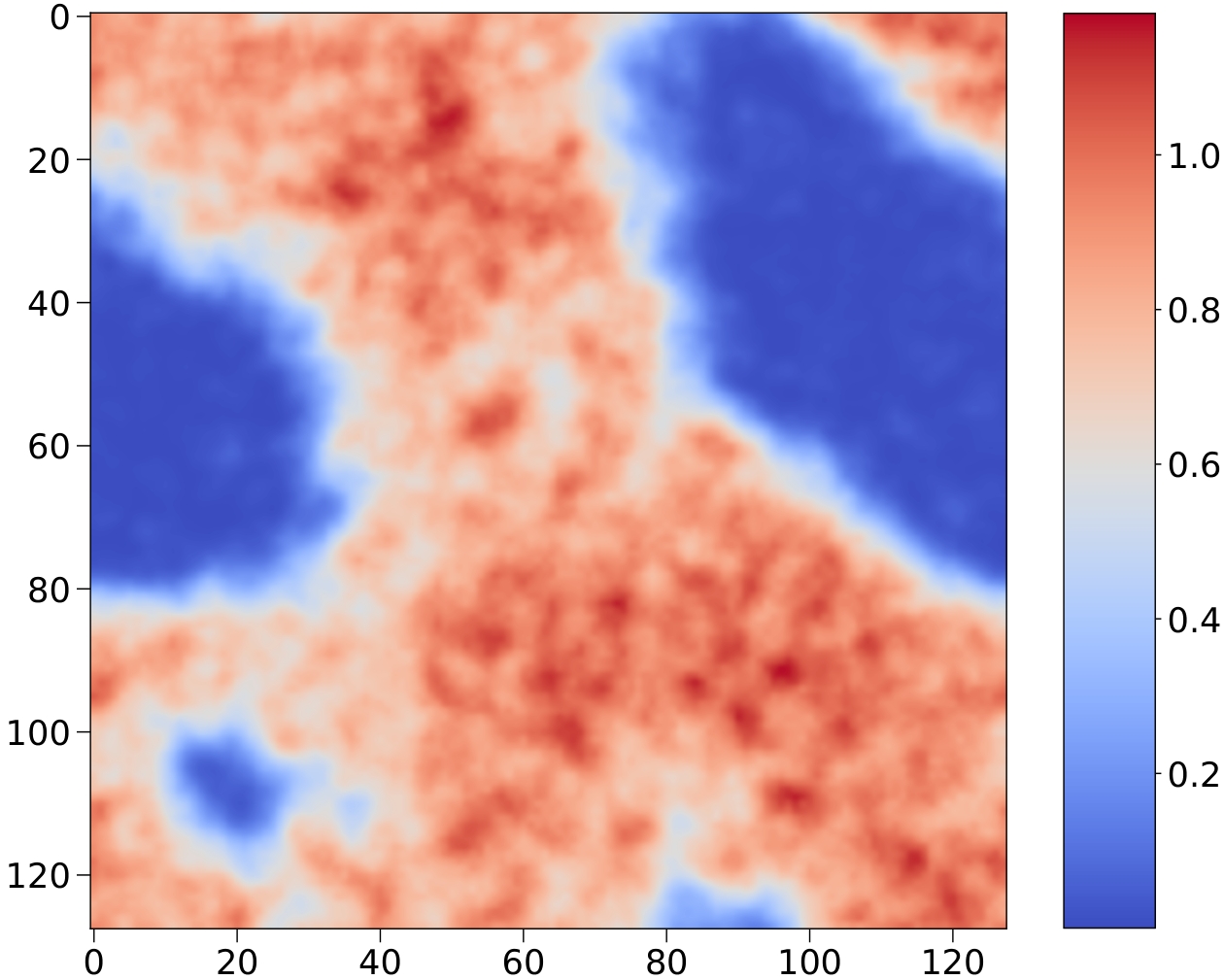}
    \includegraphics[width=0.3\linewidth]{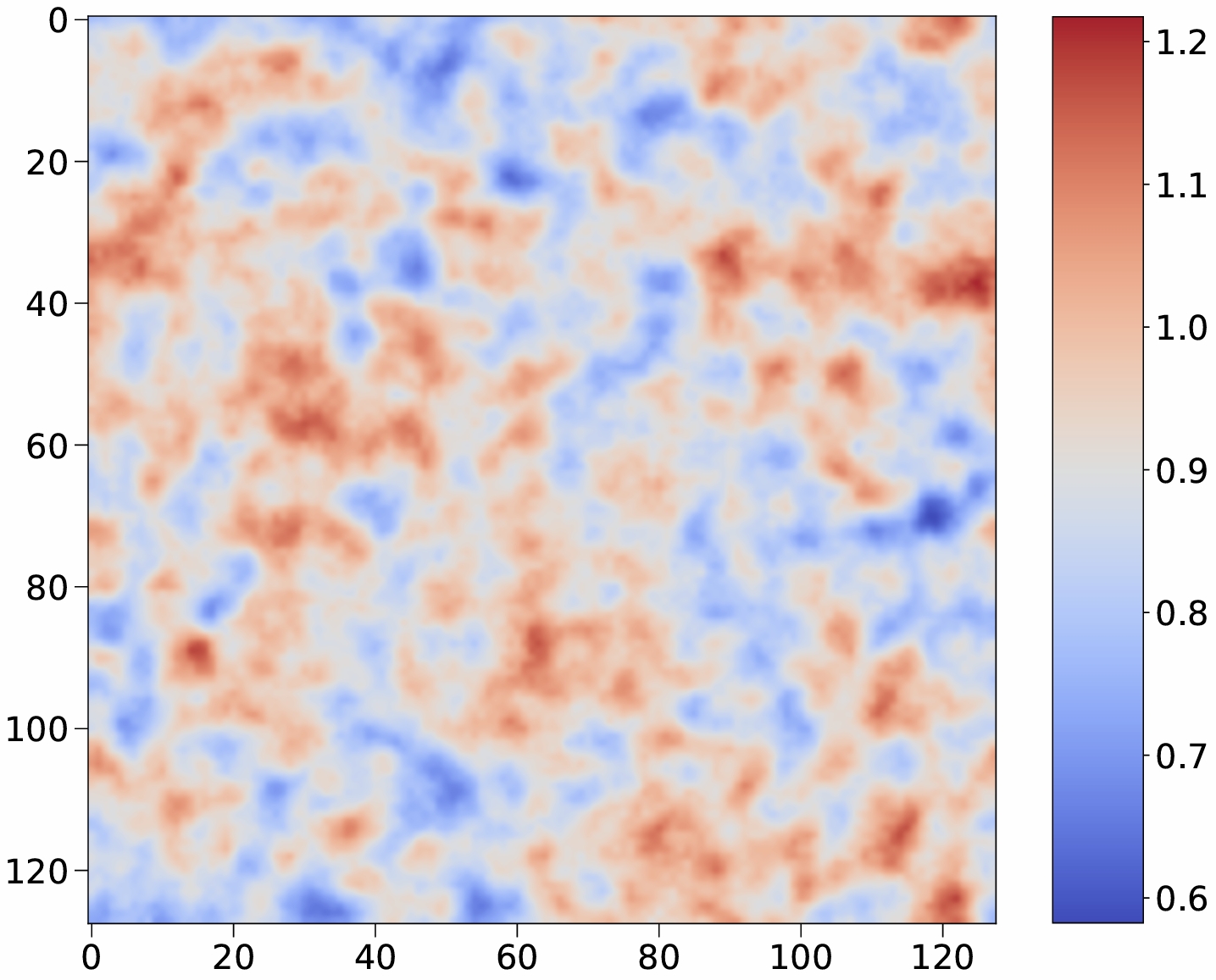}
    \includegraphics[width=0.3\linewidth]{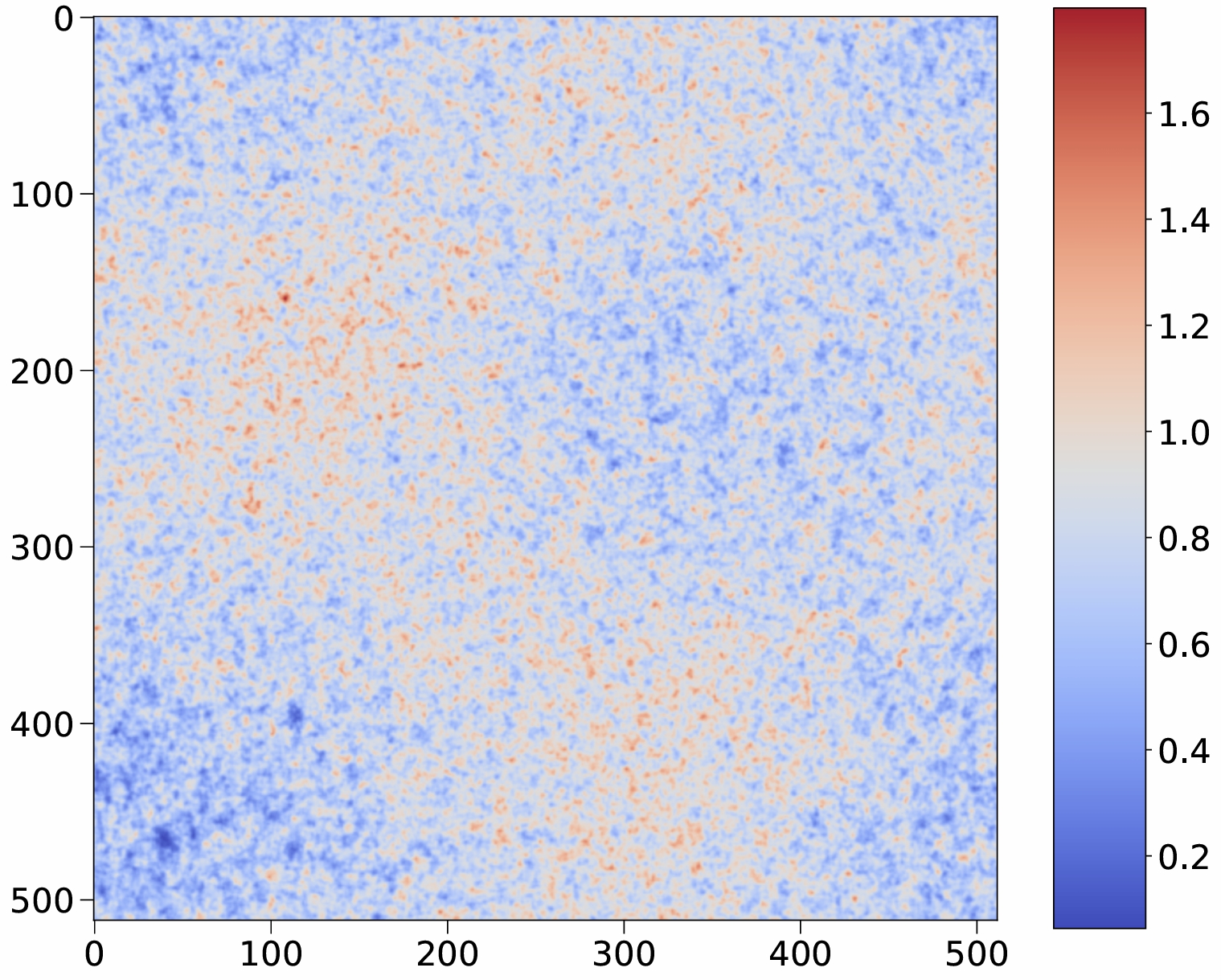}\\
    \includegraphics[width=0.3\linewidth]{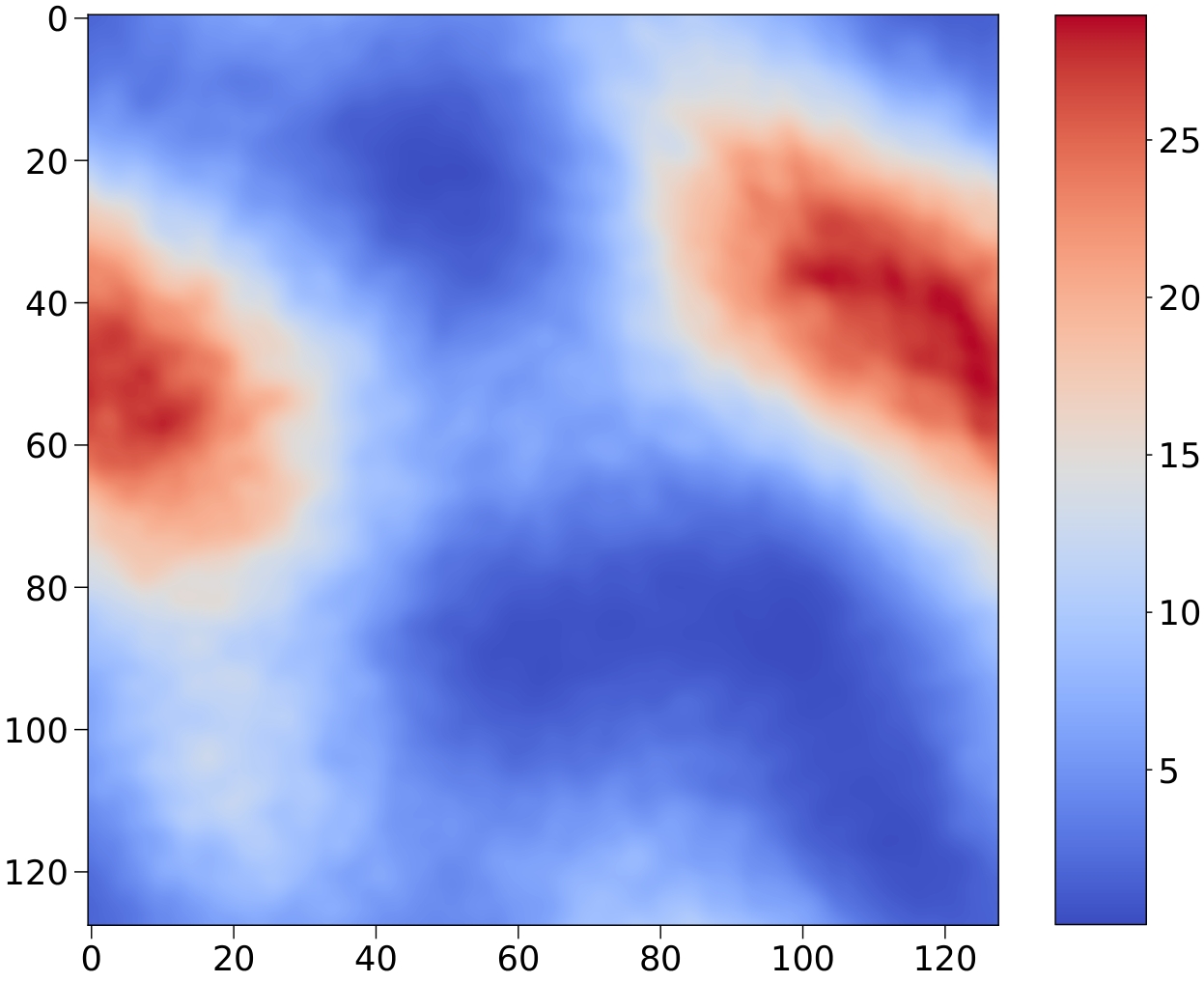}
    \includegraphics[width=0.3\linewidth]{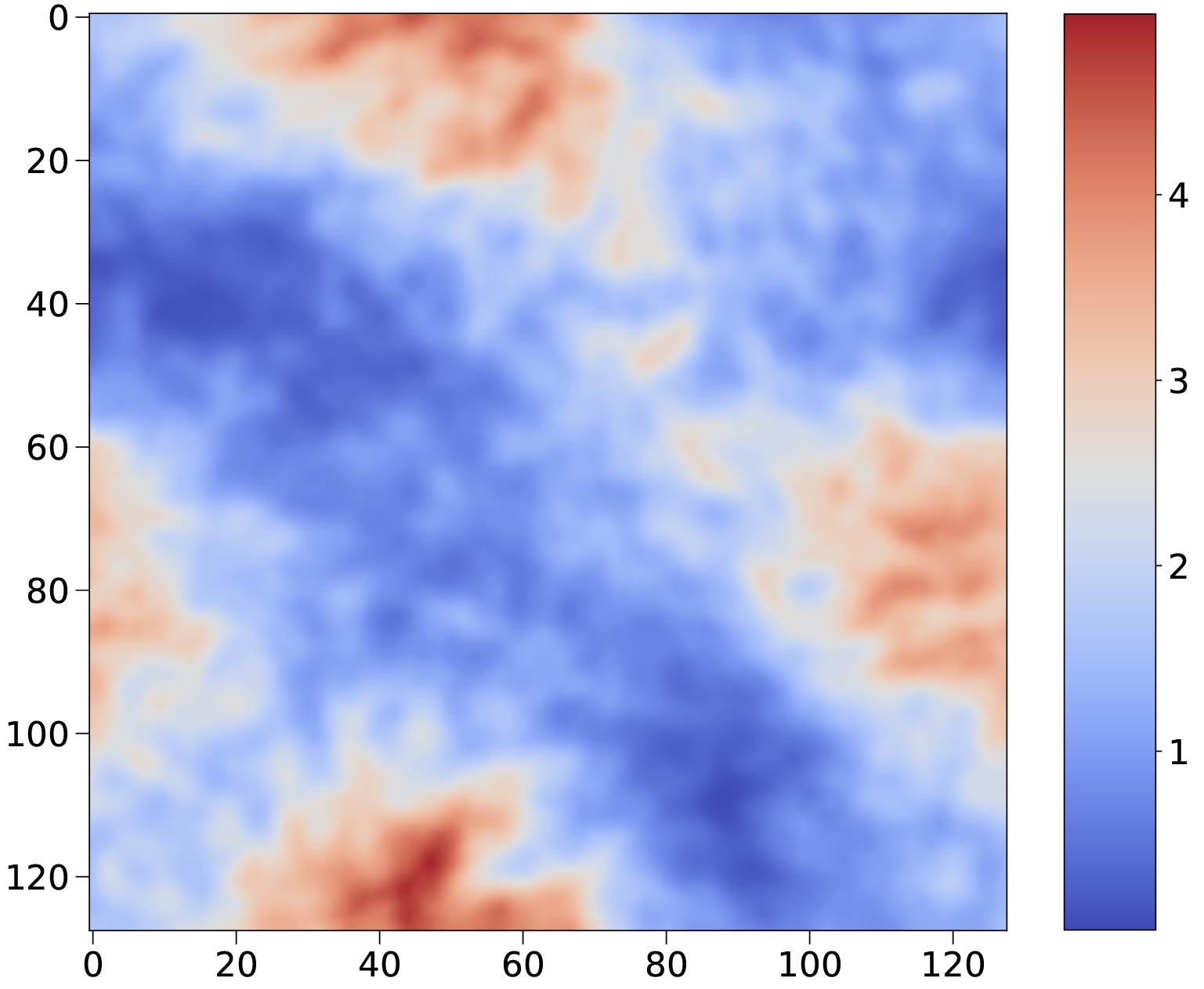}
    \includegraphics[width=0.3\linewidth]{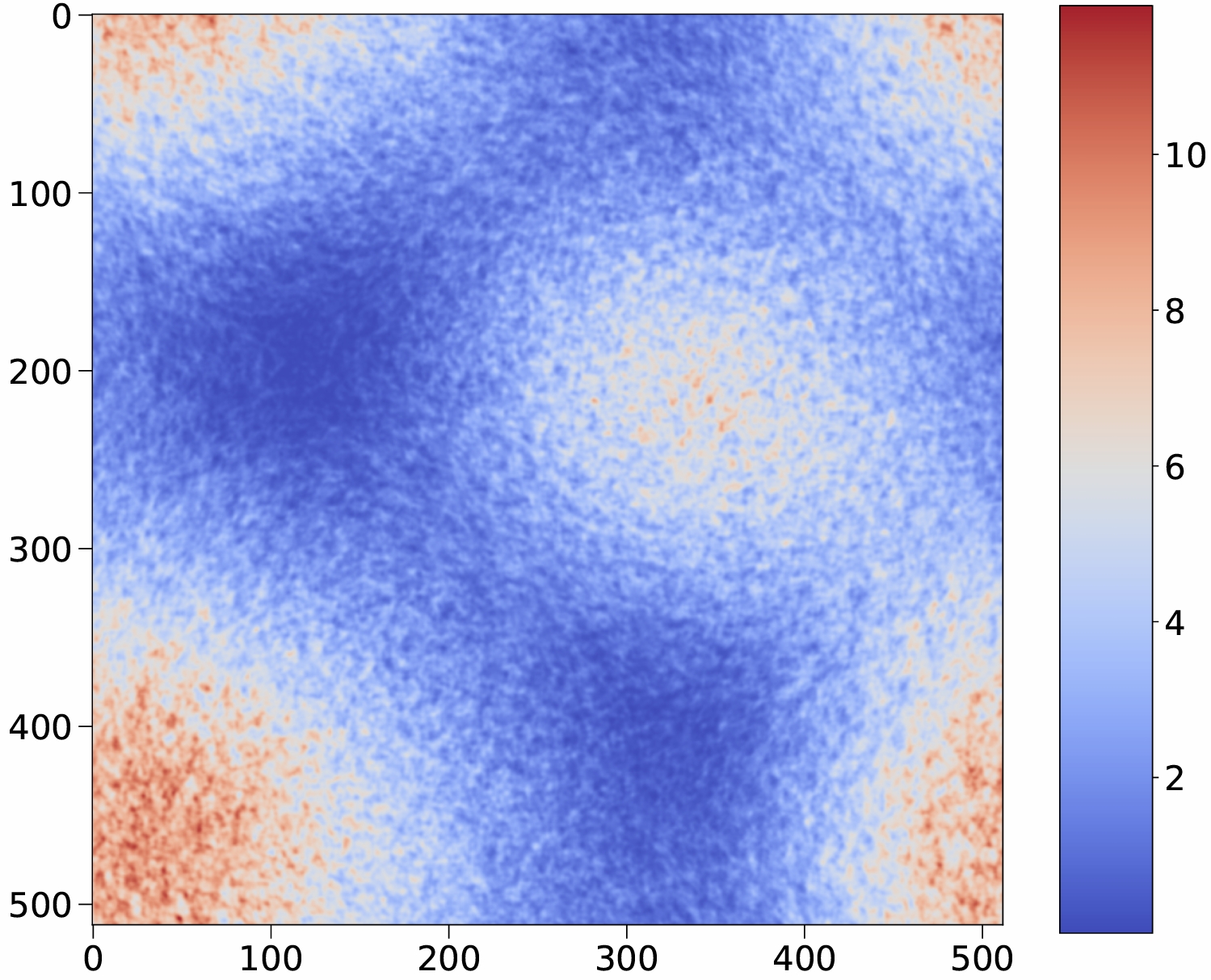}
    \caption{Slice of $|\Phi|^2$ (Top) and hypermagnetic energy (Bottom) at the end of the simulation. The hyperMF has a spectral distribution with correlation length of $\lambda_{BY}\sim R_0$ and with spectral indices of 0 (Left) and 1 (Middle), and with $\lambda_{BY}\sim R_*$, $n=0$ (right), respectively.}
    \label{fig:phiandBY}
\end{figure*}

Fig. \ref{fig:Higgs2} shows the changes of $\overline {\Phi^2}$ and $\overline {B^2} = \frac{1}{NN_t}\sum_{t,z}B^2(t,x,y,z)$ under different homogeneous hyperMFs. When $g'B_Y^\mathrm{ex}/m^2_W>3.63$, the Higgs condensate phenomenon appears.
It can be seen that $\overline {\Phi^2}$ is relatively small at the center of the vortex, which is exactly the opposite of $\overline {\Phi^2}$. It is worth mentioning that every time step of the Higgs field is involved in the averaging, while MF is only averaged every 100 steps, so the picture of $\overline {B^2}$ is rougher. We also find that the stronger the hyperMF, the denser the vortex.

When the background MF is entirely provided by the U(1) hyperMF, after the electroweak symmetry breaking, the electromagnetic field part will also undergo a phenomenon similar to the W and Z condensation. For analytical calculations and lattice simulations related to this phenomenon, see Refs.~\cite{Chernodub:2012fi, VanDoorsselaere:2012zb, Chernodub_2023}.

Slices of $|\Phi|^2$ and hypermagnetic field energy at the end of the simulation are shown in Fig.~\ref{fig:phiandBY}. The hyperMF with a spectral distribution is no longer homogeneous in space, but with some places having high energy and some places being almost zero. At the end of the simulation, because the background MF does not change with time, the distribution of hyperMF energy is similar to that at the beginning. However, it should be noted that $|\Phi|^2$ at the end tells us that the PT is only successfully and quickly completed in areas with small hyperMFs, while in places with large hyperMFs, $\Phi$ is still in the symmetric phase. Due to the random hyperMF distribution, unlike the homogeneous hyperMF with a fixed direction, the Higgs condensation phenomenon cannot be observed.

\subsection{Electroweak Sphaleron}

\begin{figure}[!htp]
    \centering
    \includegraphics[width=0.4\textwidth]{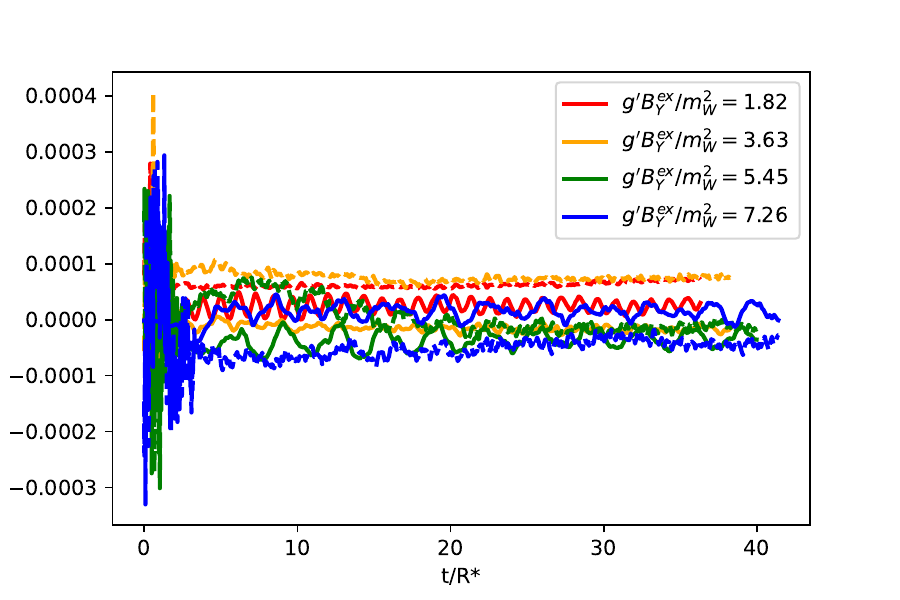}
    \includegraphics[width=0.4\textwidth]{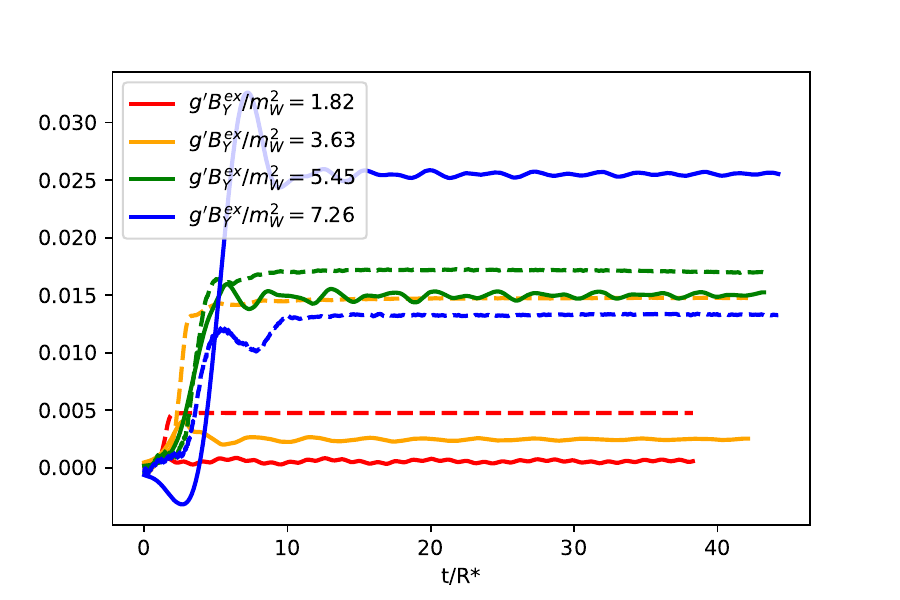}
    \caption{The variation of Chern-Simons Number $N_\mathrm{CS}$ (solid line) and Higgs winding number $N_\mathrm{H}$ (dashed line) over time under homogeneous external MFs with $h_\mathrm{factor} = 0$ (Top) and $h_\mathrm{factor} = 1$. }
    \label{fig:NW1}
\end{figure}

\begin{figure}[!htp]
    \centering
    \includegraphics[width=0.4\textwidth]{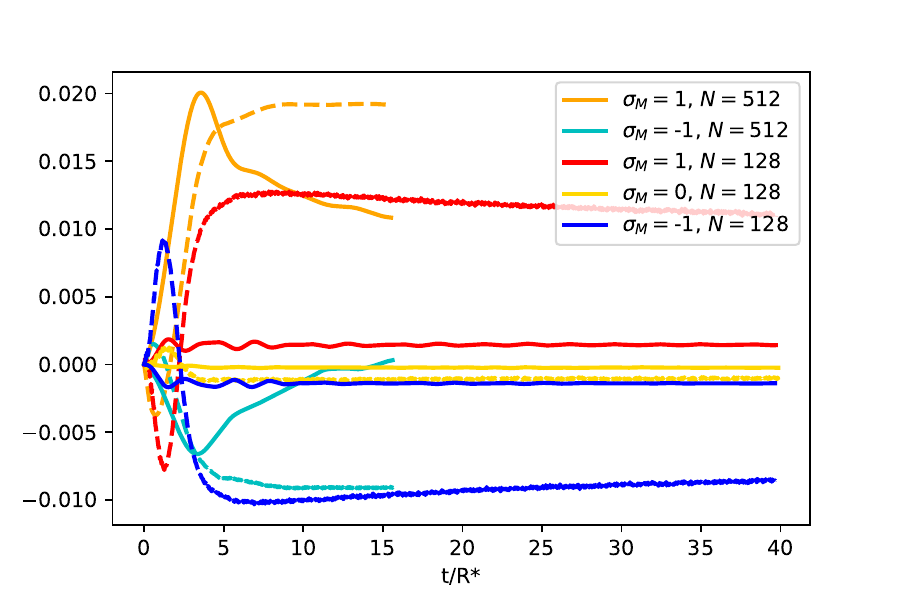}
    \includegraphics[width=0.4\textwidth]{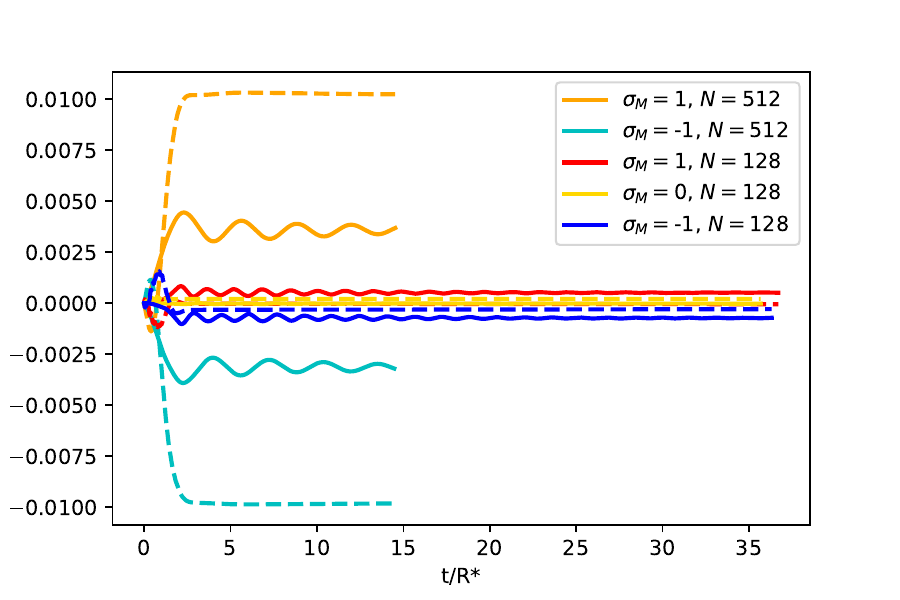}
    \caption{The variation of Chern-Simons Number $N_\mathrm{CS}$ (solid line) and Higgs winding number $N_\mathrm{H}$ (dashed line) over time under spectral-distributed external MFs with $n = 0$ (Top) and $n=1$ (Bottom). }
    \label{fig:NW2}
\end{figure}

We present the relationship between the Chern-Simons number (and the Higgs winding number) and the helicity of the MFs in Fig.~\ref{fig:NW1} and Fig.~\ref{fig:NW2}, which reflect the increase of the $N_\mathrm{CS}$ and $N_\mathrm{H}$ as the helicity of the MFs grows. According to our previous research~\cite{Di_2021}, in a first-order PT, a MF is generated due to the collision of bubbles. But when this process occurs in the context of a hyperMF, the situation will undergo some changes.
 When a PT occurs, there would be a slight increase in MF strength as the bubble nucleates. Subsequently, the bubble expands, and the MF will stabilize for a short period of time. When bubbles collide with each other, which is a very intense dynamic process, the $N_\mathrm{CS}$ rapidly increases or decreases. Moreover, the stronger the external MF, the larger the $N_\mathrm{CS}$ generated. The magnitude of the Higgs winding number $N_\mathrm{H}$ increases with the proceeding of PT when the Higgs field value increases.  After the PT is completed, the Chern-Simons number and the Higgs winding number tend to stabilize. As depicted for the two external MF scenarios, the $N_\mathrm{CS}$ grows as the strength and helicity of the hyperMF increase. Meanwhile, we observe the difference between the change of Chern-Simons number $\Delta N_\mathrm{CS}$ and the change of the Higgs winding number $\Delta N_\mathrm{H}$ during the PT. The relationship between the Chern-Simons number and the MF helicity here does not support the existence of the electroweak strings reported in Refs.~\cite{PhysRevLett.101.171302, PhysRevLett.87.251302,Vachaspati:1994ng,Barriola:1994ez}, since we didn't assume any ansatz for gauge fields.

\begin{figure}[!htp]
    \centering
    \includegraphics[width=0.4\textwidth]{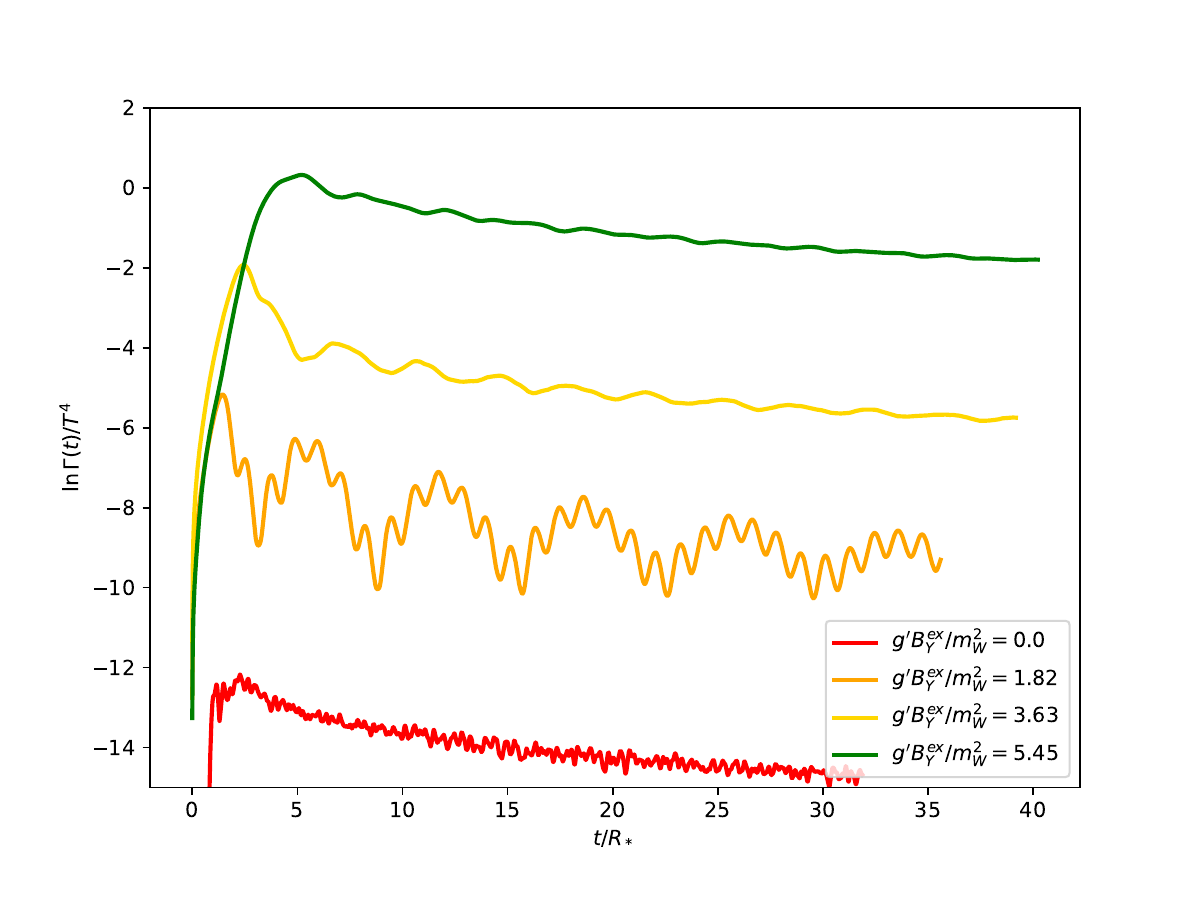}\\
    \includegraphics[width=0.4\textwidth]{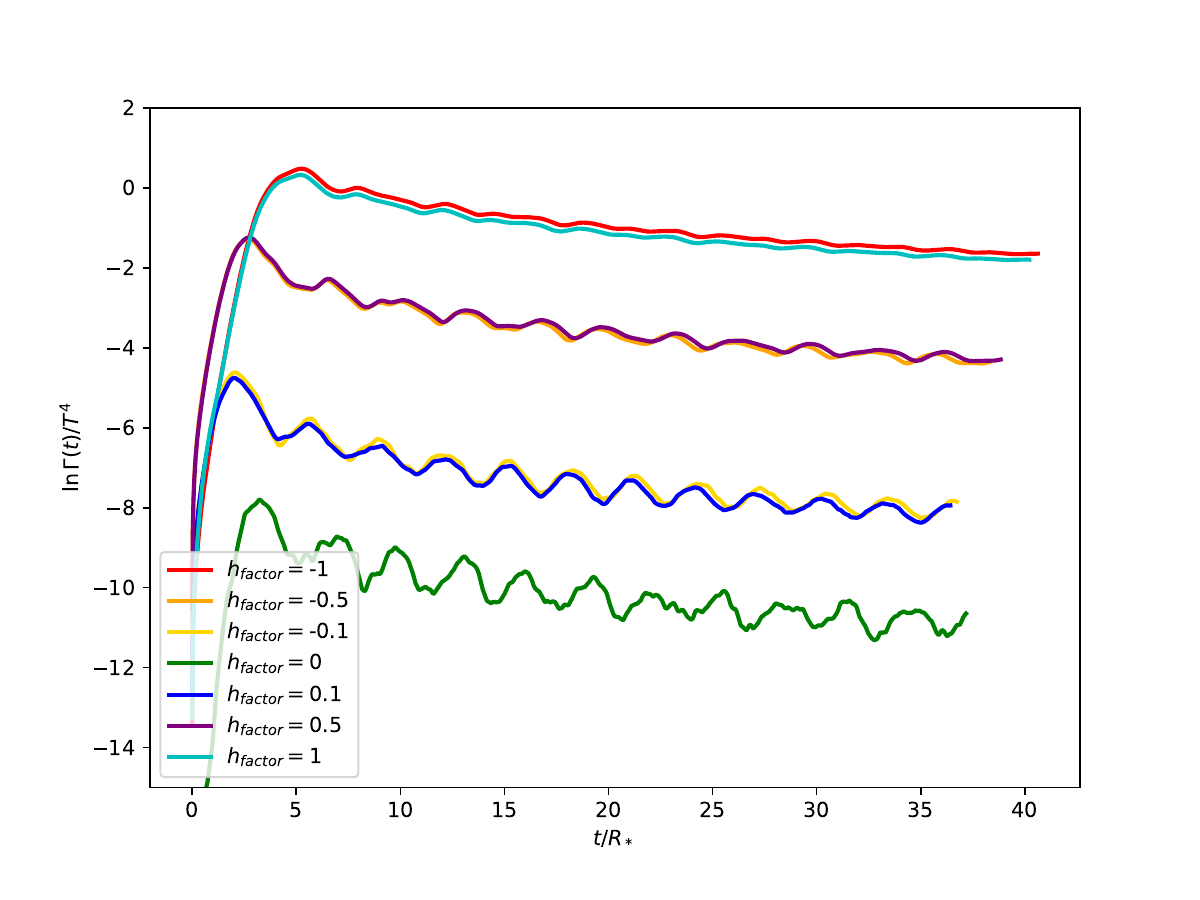}
    \caption{Top: Changes of $(\ln\Gamma(t)/T^4)$ over time under different homogeneous helical MFs with $h_\mathrm{factor} = 1$. Bottom: Changes of $(\ln\Gamma(t)/T^4)$ over time at different initial helicity with homogeneous hyperMF fixed at $g'B_Y^\mathrm{ex}/m_W^2 = 5.45$. }
    \label{fig:helGamma}
\end{figure}

We first use the diffusion rate usually adopted to describe the evolution of the Chern-Simons number closing to thermal equilibrium , which is also known as the sphaleron rate 
\begin{align}
    \Gamma(t) = \frac{\langle (\Delta N_\mathrm{CS}(t))^2\rangle}{Vt}\label{eq:Gamma}
\end{align}
to measure the intensity of the change of the volume averaged $N_\mathrm{CS}$. Angle brackets indicate ensemble averaging, which is the average of multiple runs.


\begin{figure}[!htp]
    \centering
    \includegraphics[scale=0.4]{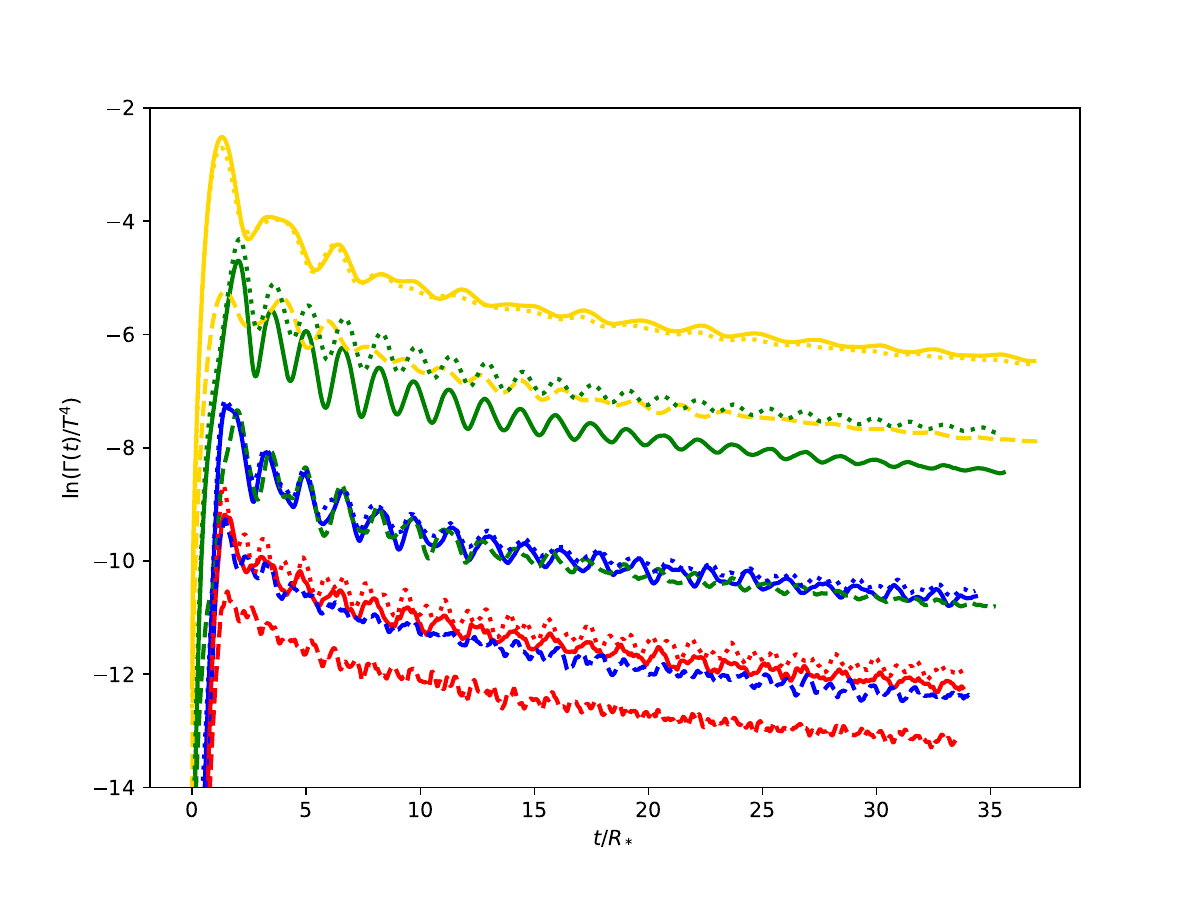}
    \includegraphics[scale=0.4]{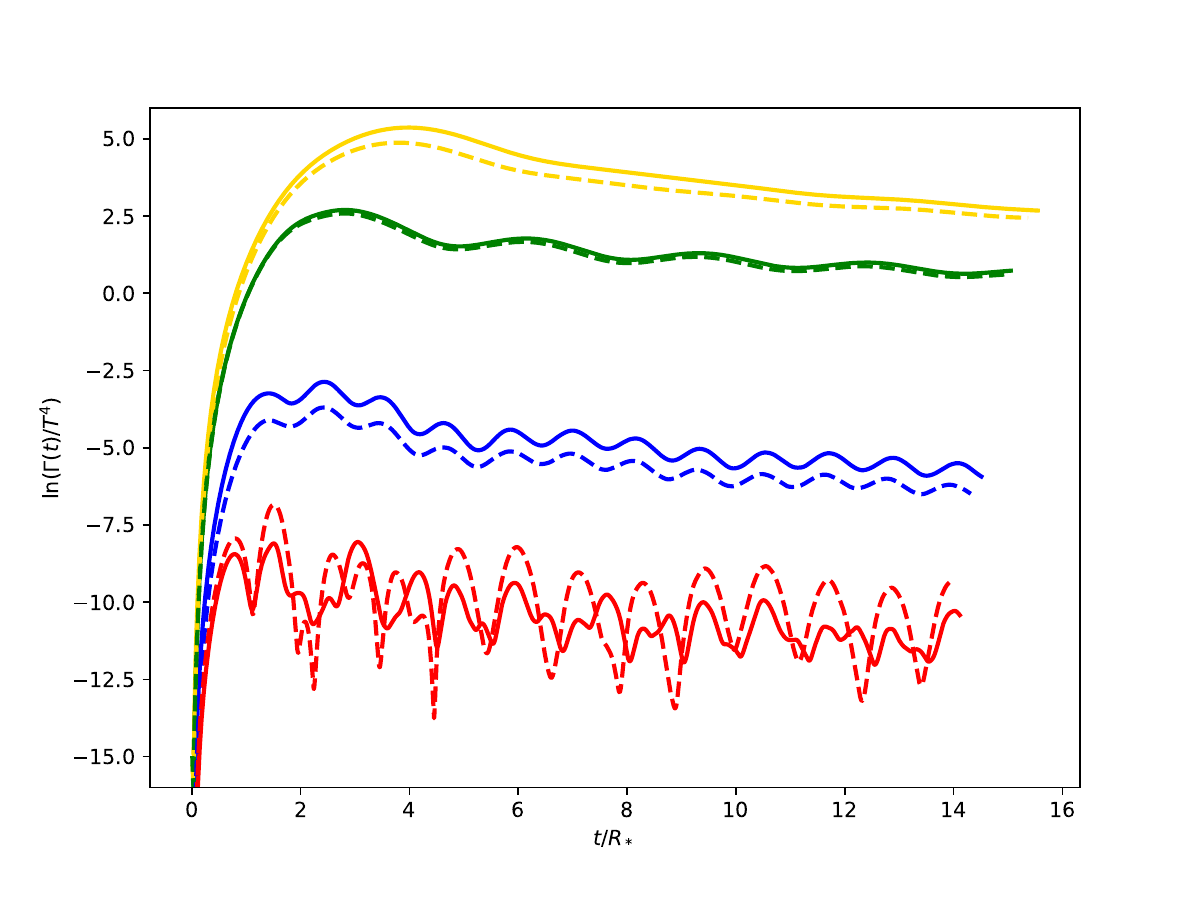}
       \caption{
Evolution of $\ln\Gamma(t)/T^4$ over time under different index $n$ of the spectrum distributed hyperMF and correlation length ($\lambda_B \sim R_0$ on the top and $\lambda_B \sim R_*$ on the bottom.). $n = 0,1,2,3$ corresponds to yellow, green, blue, and red lines, respectively.  The solid line indicates $\sigma_\mathrm{M}=1$, the dashed line indicates $\sigma_\mathrm{M}=-1$, and the dotted line indicates $\sigma_\mathrm{M}=0$, respectively.  }
    \label{fig:SBGamma1}
\end{figure}

\begin{figure}[!htp]
    \centering
    \includegraphics[scale=0.5]{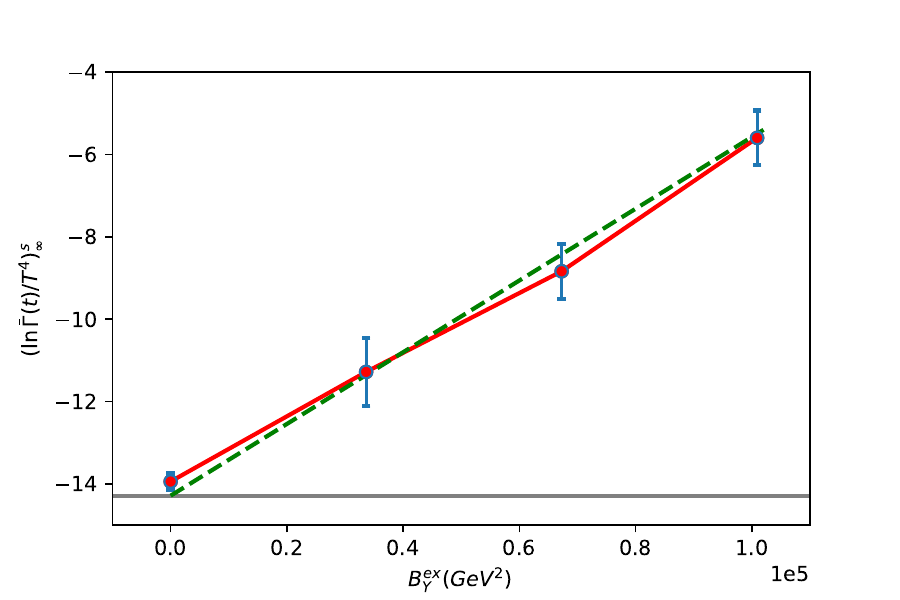}
    \includegraphics[scale=0.5]{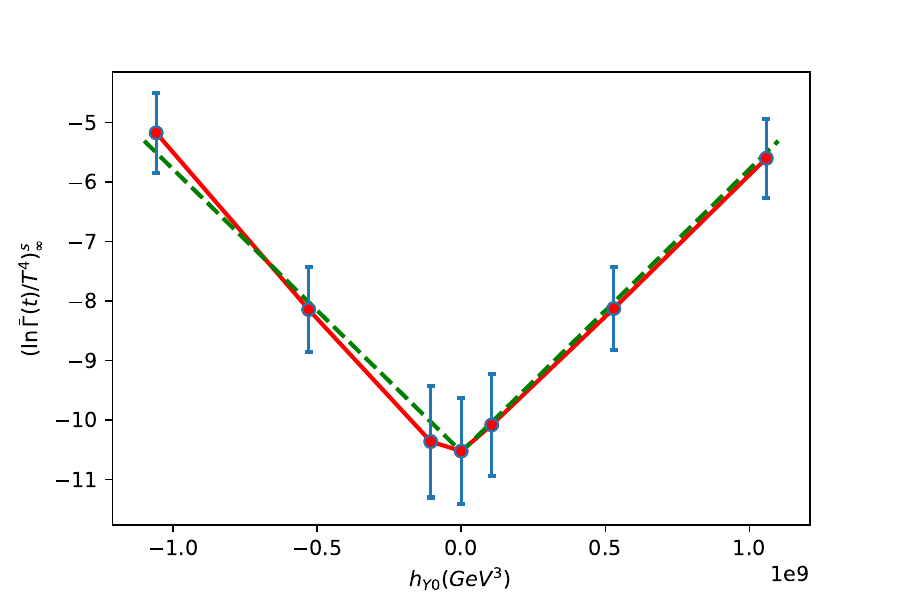}
    \caption{Top: Changes of $(\ln\,\bar\Gamma(t)/T^4)^\mathrm{s}_\infty$ over different homogeneous, helical MFs. The gray line represents the value of the sphaleron rate in the symmetric phase $\Gamma_\mathrm{sym}/T^4=6.23\times10^{-7}$~\cite{PhysRevD.107.073006}. Bottom: Changes of $(\ln\,\bar\Gamma(t)/T^4)^\mathrm{s}_\infty$ over different initial helicity with homogeneous hyperMF fixed at $g'B_Y/m_W^2 = 5.45$. The fitting result of Eq. \eqref{eq:fitGamma} is represented by the green dashed line. }
    \label{fig:helvsGamma}
\end{figure}

We first study the EW sphaleron rate behavior under the impact of the homogeneous helical MF. We find that a stronger helical hyperMF background will bring about a more drastic change in $N_\mathrm{CS}$, which is reflected by the behavior of $\ln\Gamma(t)/T^4$. As shown in Fig.~\ref{fig:helGamma},  the increase of hyperMF strength will significantly elevate $\Gamma(t)$, that is, the rate of baryon production is faster. If the hyperMF is fixed and the helicity is increased, the rate of baryon production will also increase, but the effect is not as obvious as that brought by increasing the MF strength. The change of time-averaged sphaleron rate $\Gamma(t)$ with time under different spectral indices is shown in Fig.~\ref{fig:SBGamma1}. It can be seen that $\ln\Gamma(t)/T^4$ with the same color represents the same spectral index and similar hyperMF strength. As the MF energy increases, the sphaleron rate also increases. 

To properly incorporate the PT dynamics, We further adopt the following form of the sphaleron rate \cite{Tranberg:2003gi}
\begin{align}
    \Gamma(t) = \frac1V\frac{\mathrm{d}}{\mathrm{d}t}
    [\langle (\Delta N_\mathrm{CS}(t))^2\rangle - \langle \Delta N_\mathrm{CS}(t)\rangle^2], 
\end{align}
then 
based on this $\Gamma(t)$, we define
\begin{align}
    \bar\Gamma(t) = \frac{1}{t-t_0}\int_{t_0}^t\mathrm{d}t'\, \Gamma(t'),
\end{align}
to describe the time-averaged sphaleron rate. We here calculate its time average for different hyperMF when it is stable after $t/R*>2.67$, denoted as $(\ln\bar\Gamma(t)/T^4)^\mathrm{s}$ where s stands for stable, as shown in Fig. \ref{fig:helvsGamma}. As the MF and helicity increase, $(\ln\bar\Gamma(t)/T^4)^\mathrm{s}$ also increases. Note that our initial $Y_\mathrm{ex}$ and hyperMF settings for the helical situation are as shown in Section~\ref {App:setup} and Appendix~\ref {sec:hyMFlat}, so the initial helicity is $h_{Y0} = h_\mathrm{factor}L(B^\mathrm{ex}_Y)^2$. When the hyperMF strength is changed, the helicity $h_{Y0}$ will change simultaneously. Therefore, we can fit the dependence of $(\ln\bar\Gamma(t)/T^4)^\mathrm{s}$ on MF and helicity:
\begin{align}
    \left(\ln\frac{\bar\Gamma(t)}{T^4}\right)^\mathrm{s}_\infty = 
    (0.78 |h_\mathrm{factor}| + 0.58)\left(\frac{B^\mathrm{ex}_Y}{T^2}\right) + \ln\frac{\Gamma_\mathrm{sym}}{T^4}\;,\label{eq:fitGamma}
\end{align}
which is shown in Fig.~\ref{fig:helvsGamma} by the green dashed line. The subscript $\infty$ represents that the external hyperMF under study has an infinite correlation length. $\Gamma_\mathrm{sym}/T^4=6.23\times10^{-7}$ represents the value of the sphaleron rate in the symmetric phase~\cite{PhysRevD.107.073006}, which is showed by the gray line in Fig. \ref{fig:helvsGamma}. Here, $T$ indicates the PT temperature. We can see that the sphaleron rate can grow by many orders with the increase of the strength and helicity of the external hyperMF.


\begin{figure}[!htp]
    \centering
    \includegraphics[scale=0.5]{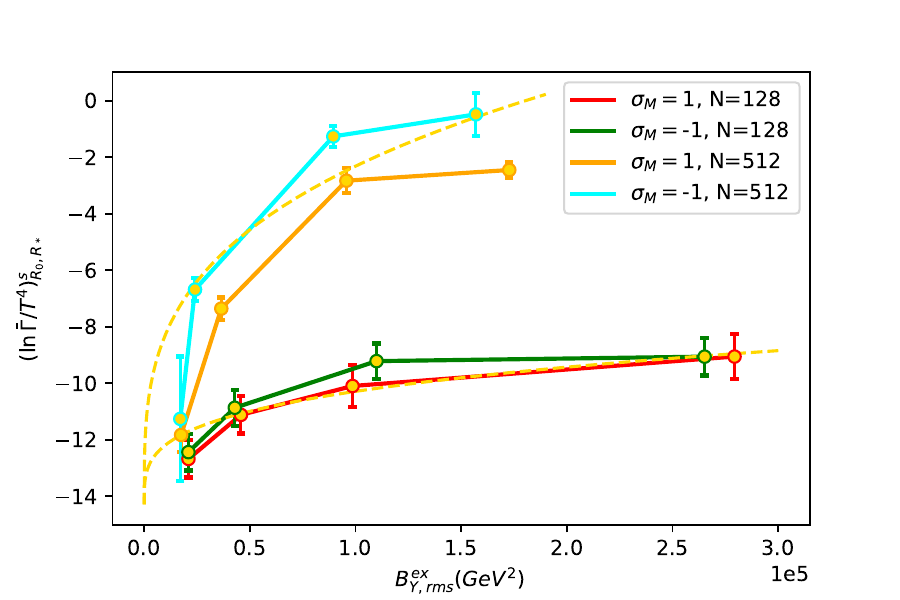}
    \caption{Changes of $(\ln\bar\Gamma(t)/T^4)^\mathrm{s}_{R_0,R_*}$ over different hypermagnetic energy with $\sigma_\mathrm{M}=\pm1$ and correlation length of $\lambda_B\sim R_0$ and $\lambda_B\gtrsim R_*$. The yellow dashed line represents the fitting result of Eq. \eqref{eq:fitGammaSB} and Eq. \eqref{eq:fitGammaSB512}. }
    \label{fig:SBGamma}
\end{figure}

Similarly, we extract $(\ln\bar\Gamma(t)/T^4)^\mathrm{s}$ and plot the variation of electroweak Sphaleron rate versus hyperMF strength in Fig.~\ref{fig:SBGamma} and use the yellow dashed line to show the following fitting results: 
\begin{align}
    \left(\ln\frac{\bar\Gamma(t)}{T^4}\right)^\mathrm{s}_{R_0} &= 
    2.38|\sigma_\mathrm{M}|\left(\frac{B_{Y\mathrm{,rms}}}{T^2}\right)^{0.28} + \ln\frac{\Gamma_\mathrm{sym}}{T^4}\;,\label{eq:fitGammaSB}\\
    \left(\ln\frac{\bar\Gamma(t)}{T^4}\right)^\mathrm{s}_{R_*} &= 
    6.86|\sigma_\mathrm{M}|\left(\frac{B_{Y\mathrm{,rms}}}{T^2}\right)^{0.30} + \ln\frac{\Gamma_\mathrm{sym}}{T^4}\;.\label{eq:fitGammaSB512}
\end{align}
As indicated by the figure and the above two formulas, the electroweak sphaleron rate increases by several orders when the correlation length increases from $R_0$ to $R_*$.


\subsection{Baryon and Lepton asymmetry}
\label{sec:cpla}


\begin{figure}[!htp]
    \centering
    \includegraphics[scale=0.4]{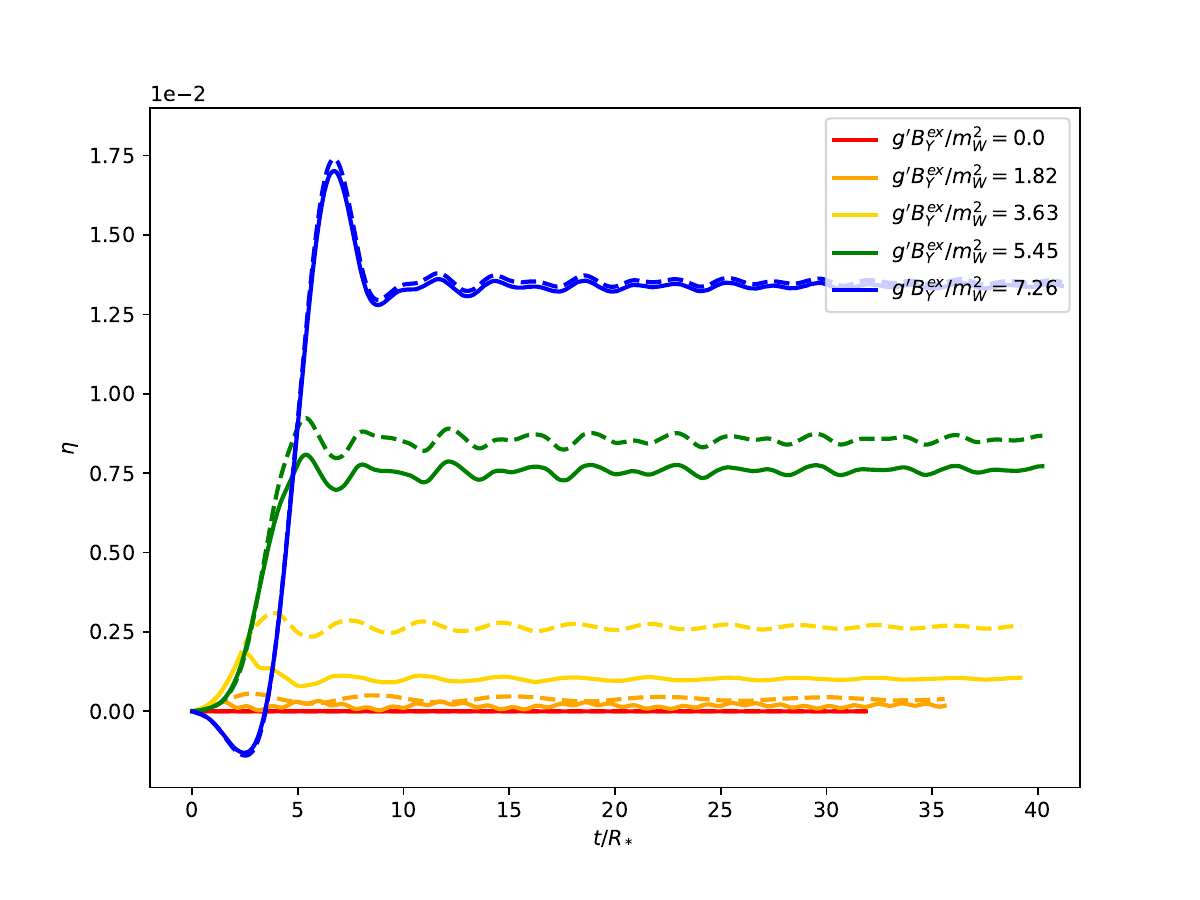}
    \includegraphics[scale=0.4]{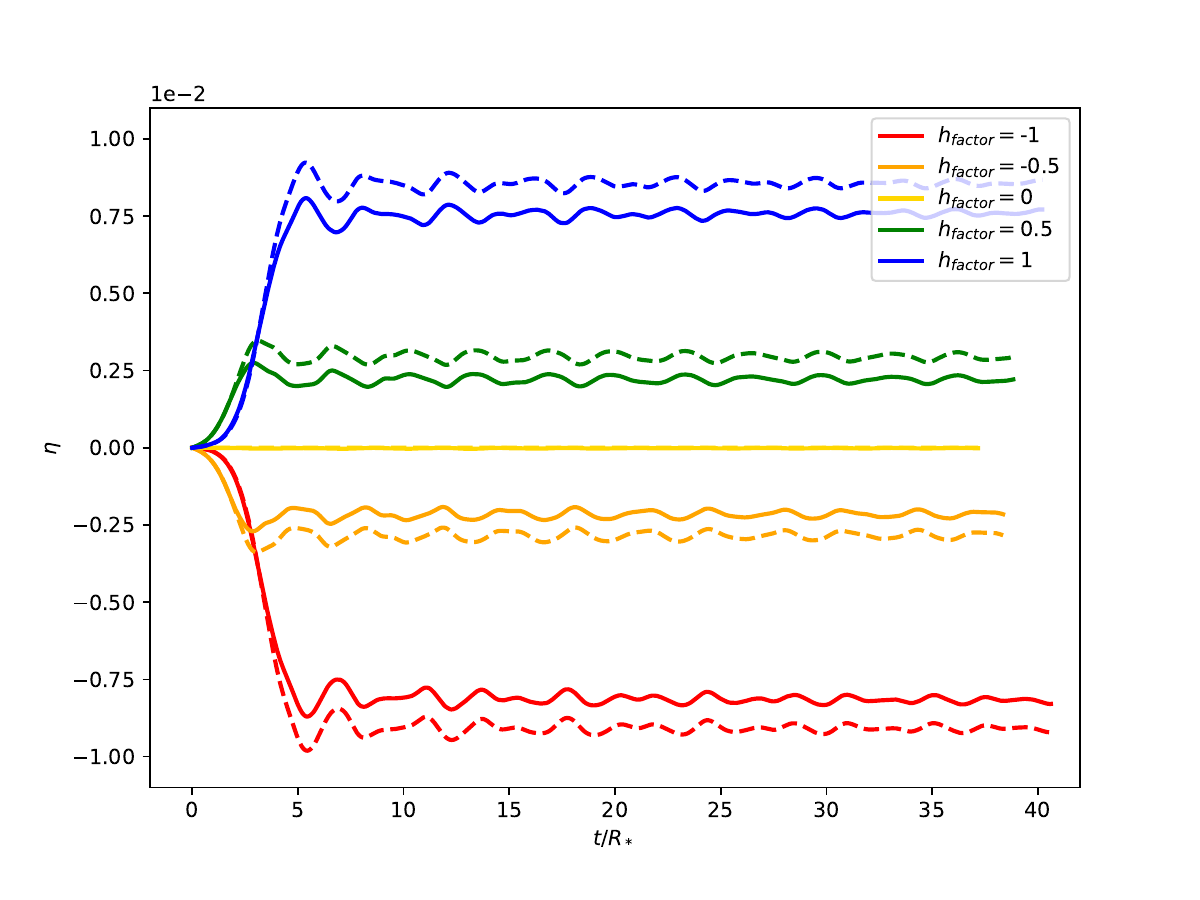}
    \caption{Top: The variation of $\eta_\mathrm{B}$ (solid line) and $\eta_\mathrm{B}^Y$ (dashed line) over time under different homogeneous helical hyperMF strengths with $h_\mathrm{factor}=1$. Bottom: The variation of $\eta_\mathrm{B}$ (solid line) and $\eta^Y_\mathrm{B}$ (dashed lind) over time under different $h_\mathrm{factor}$ with $g'B_Y^\mathrm{ex}/m_W^2=5.45$.}
    \label{fig:etavst}
\end{figure}

\begin{figure}[!htp]
    \centering
    \includegraphics[scale=0.4]{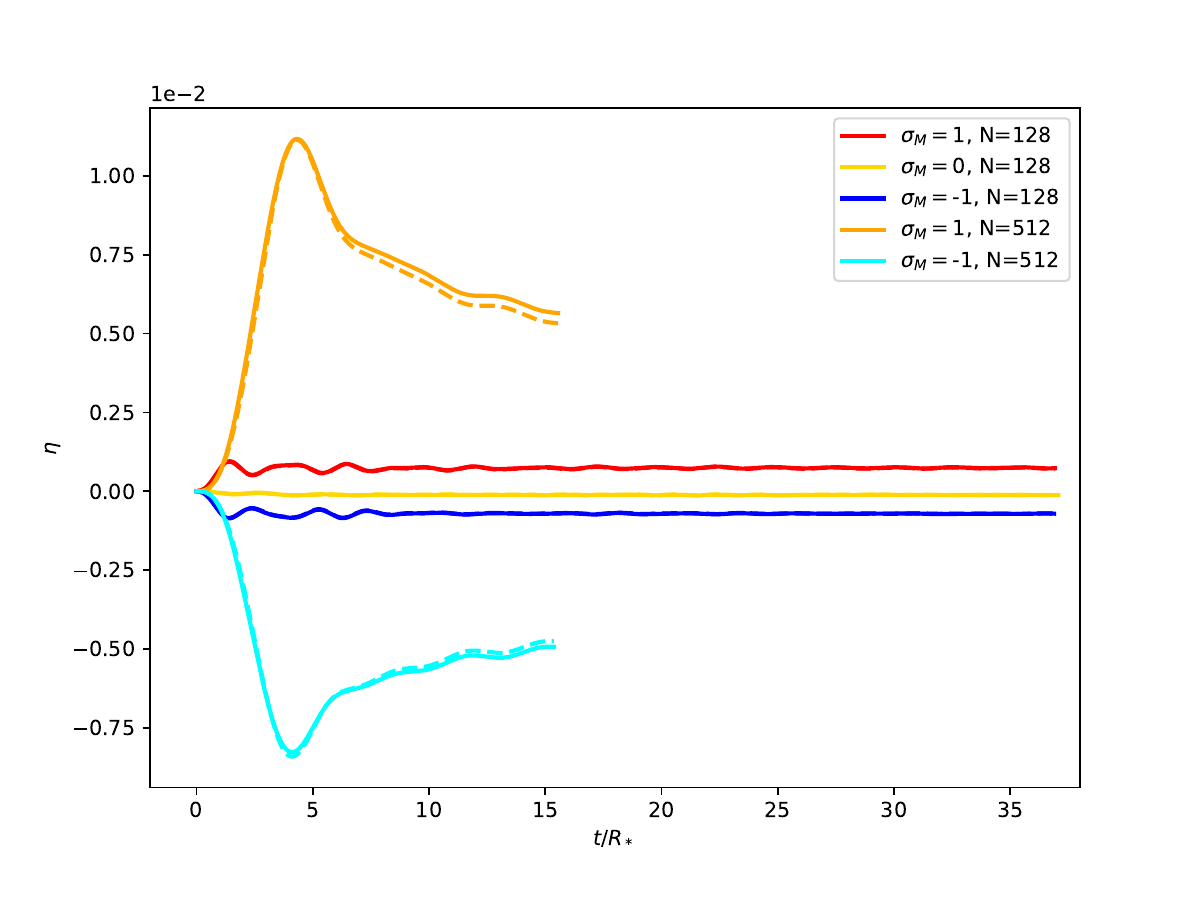}
    \includegraphics[scale=0.4]{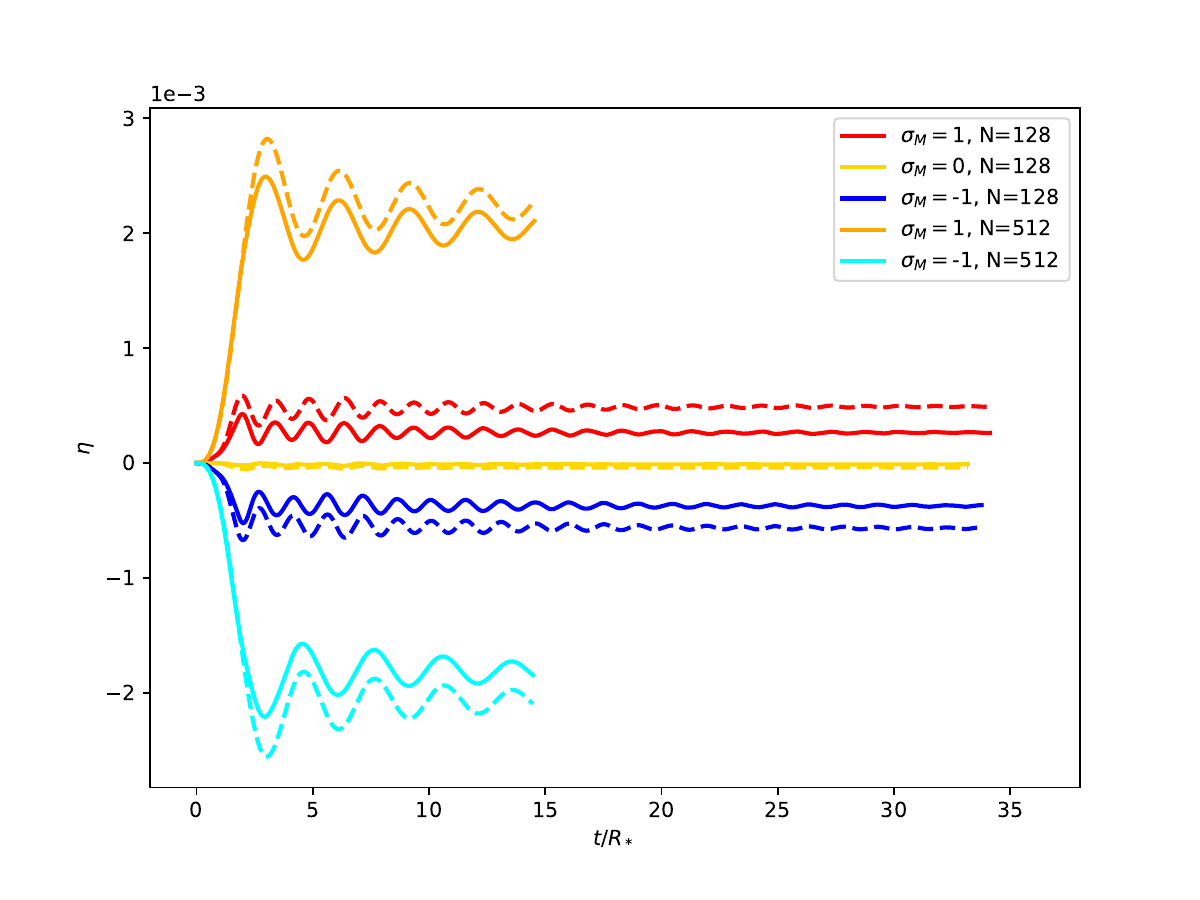}
    \caption{The variation of $\eta_\mathrm{B}$ (solid line) and $\eta_\mathrm{B}^Y$ (dashed line) over time under different spectrum distributed hyperMF's $\sigma_\mathrm{M}$ with the spectral index of $n=0$ (Top) and $n=1$ (Bottom).}
    \label{fig:SBetavst}
\end{figure}

We now investigate the time variation of the total baryon asymmetry ($\eta_\mathrm{B}$) through Eq.~(\ref{eq:nbs}) and the one induced by the helical hyperMF effect ($\eta_\mathrm{B}^Y$) through the chiral anomaly~\cite{PhysRevLett.108.031301}, i.e., Eq.~\eqref{etabY}. The changes of $\eta_\mathrm{B}$ and $\eta_\mathrm{B}^Y$ over time are shown in Fig.~\ref{fig:etavst} and Fig. \ref{fig:SBetavst}. When PT occurs, both $\eta_\mathrm{B}$ and $\eta_\mathrm{B}^Y$ increase or decrease simultaneously, depending on the sign of their helicity, and tend to stabilize upon completion of the PT. 
In the case of a spectral distribution hyperMF, we present $\eta_\mathrm{B}$ and $\eta_\mathrm{B}^Y$ for different helical hyperMF spectra. 
It can be seen that $\eta_\mathrm{B}^Y\gtrsim \eta_\mathrm{B}$, and $\eta_\mathrm{B}^Y$ occupies the dominant part of $\eta_\mathrm{B}$. For both the homogeneous and spectral distribution of the helical hyperMF scenarios, we have that the generation of the baryon asymmetry increases as the helicity and magnitude of the hyperMF increase.

\subsection{Remark on the MF}
\label{sec:rhymf}

\begin{figure*}[!htp]
    \centering
\includegraphics[width=0.8\linewidth]{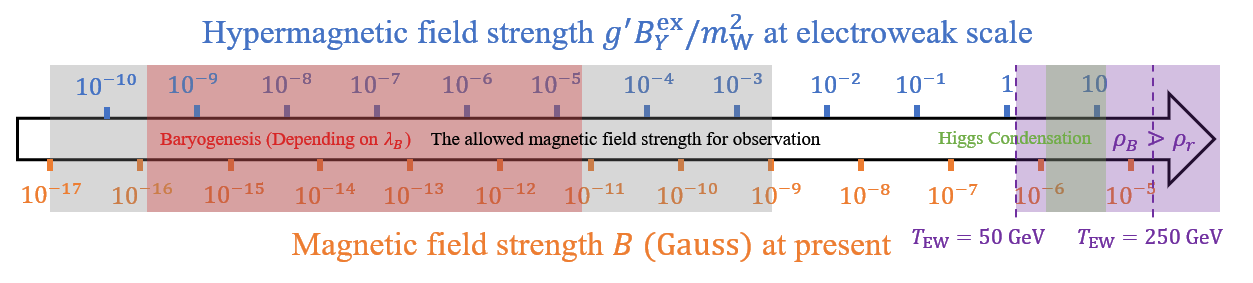}
    \caption{Schematic diagram of the range of the cosmic MF. The observations of CMB and blazar, respectively, limit the upper and lower limits of MF strength \cite{Fujita:2016igl, Durrer_2013}. The radiation energy density in the early universe limited the upper limit of the MF strength at that time. The range of MF strength that can achieve correct matter-antimatter asymmetry obtained in Ref.~\cite{Di:2024gsl} is also shown in the figure.}
    \label{fig:Brange}
\end{figure*}

Consider MFs with finite correlation lengths and under fully helical conditions, by using the magnetohydrodynamic equation, we can obtain the relationship between the MF during the electroweak PT epoch and the magnetic field at present. The (comoving) correlation scale evolves as $\lambda_\mathrm{rec}/\lambda_\mathrm{EW} = (a_\mathrm{rec}/a_\mathrm{EW}) (\tau_\mathrm{rec}/\tau_\mathrm{EW})^{2/3}$, and the MF strength evolves as $B_\mathrm{rec}/B_\mathrm{EW} = (a_\mathrm{rec}/a_\mathrm{EW})^{-2} (\tau_\mathrm{rec}/\tau_\mathrm{EW})^{-1/3}$~ \cite{Kahniashvili_Brandenburg_Tevzadze_2016, Brandenburg_Durrer_Huang_Kahniashvili_Mandal_Mukohyama_2020, Brandenburg:2021bfx}, where $\tau$ denotes conformal time, and the subscripts ``rec" and ``EW" refer to the recombination epoch and the electroweak PT epoch, respectively. If the MF is non-helical, then the MF will follow the following relationship: $\lambda_\mathrm{rec}/\lambda_\mathrm{EW} = (a_\mathrm{rec}/a_\mathrm{EW}) (\tau_\mathrm{rec}/\tau_\mathrm{EW})^{1/2}$, and the MF strength evolves as $B_\mathrm{rec}/B_\mathrm{EW} = (a_\mathrm{rec}/a_\mathrm{EW})^{-2} (\tau_\mathrm{rec}/\tau_\mathrm{EW})^{-1}$ \cite{Kahniashvili_Brandenburg_Tevzadze_2016}. After the recombination epoch, the density of charged particles in the universe sharply decreased, and the MF was considered to no longer evolve, but only passively ``diluted" with the expansion of the universe as $B\propto a^{-2}$. These are the two types of magnetic field strength conversion relationships shown in Fig. \ref{fig:Brange}. Fig.~\ref{fig:Brange} also shows the range of MF strength (depending on its correlation length) required to achieve the correct matter-antimatter asymmetry obtained in Ref.~\cite{Di:2024gsl}, as well as the MF required for the Higgs condensate in this work.
According to current astronomical observations, the upper limit of MF strength in the universe today is determined by CMB, which is approximately $10^{-9}$ Gauss on CMB scale $\lambda_B\gtrsim1$ Mpc \cite{refId0}, and the lower limit is set by observations of blazars as \cite{Taylor_2011}
\begin{align}
    B\gtrsim10^{-17}\ \mathrm{Gauss}
    \begin{cases}
        (\lambda_B/1\ \mathrm{Mpc})^{-1/2}, &\lambda_B<1\ \mathrm{Mpc}\;,\\
        1, &\lambda_B>1\ \mathrm{Mpc}\;.
    \end{cases}
\end{align}
During the electroweak PT, the universe is dominated by radiation, and the MF energy density should not exceed that of the radiation energy density:
\begin{align}
    \rho_r = \frac{\pi^2}{30}g_{*S}T^4, 
\end{align}
where $g_{*S}=106.75$ is the effective number of degrees of freedom in entropy before the PT. If we take the phase transition temperature as $T_\mathrm{EW} = 125$ GeV, the upper limit of (hyper)MF strength is $g'B_Y/m^2_\mathrm{W} = 7.07$. In our simulation, for the MF with spectral distribution, $n<1$ can cause the MF strength to exceed this limit. 
The above limitations are shown in Fig.~\ref{fig:Brange}. 

\section{Conclusion and discussion }\label{sec:Con}
In this work, we numerically investigated the first-order electroweak PT in the background of primordial MF through 3D lattice simulations. We investigated various effects of the primordial MF on the first-order electroweak PT, encompassing PT speed, Ambj\o rn-Oleson condensation, sphaleron rate, and Baryon and Lepton asymmetry generations. 
Firstly, we found that an external MF slows down the speed of PT. This is because the expansion speed of vacuum bubbles in a MF slows down, which also leads to the generation of more bubbles in a stronger MF. If the external MF is helical, the PT speed will slow down slightly further. Secondly, when the MF exceeds the first critical value, the  Ambj\o rn-Oleson condensation phenomenon occurs, where the Higgs field is arranged into a regular hexagonal structure due to the influence of the SU(2)$_\mathrm{L}$ gauge field. As the external MF increases, this hexagonal structure becomes increasingly dense. This phenomenon applies to both helical and non-helical external MFs. In the case of a spectrum distribution MF, since the MF does not have a definite direction, we cannot observe the Ambj\o rn-Oleson condensation phenomenon. In regions where the MF is very strong, the electroweak PT does not occur at all, and the electroweak symmetry remains (or is restored). After symmetry breaking, the SU(2)$_\mathrm{L}$ gauge field breaks down into the U(1)$_\mathrm{em}$ electromagnetic field, so the MF also exhibits a similar hexagonal structure.

Sphaleron rate $\Gamma$ is an important physical quantity that characterizes the rate of baryon number production. We studied the changes in $\Gamma$ under different preexisting MF strengths, helicities, and correlation lengths. We found that as the MF increases and the helicity increases, $\Gamma$ will also increase. The correlation scale mainly affects the power-law dependence of $\Gamma$ on external MF. We detailed here the mechanism to generate the baryon asymmetry through the hyperMF, where the MF produced during the PT ensures the time variation of hyperMF helicity, which drives baryon asymmetry through the chiral anomaly. For the relationship between baryon asymmetry and the properties of the hyperMF, and its detectability at present, we refer to Ref.~\cite{Di:2024gsl}.

\section{Acknowledgements}

 L.B. is supported by the National Key Research and Development Program of China under Grant No. 2021YFC2203004, and by the National Natural Science Foundation of China (NSFC) under Grants Nos. 12322505, 12347101. L.B. also acknowledges Chongqing Talents: Exceptional Young Talents Project No. cstc2024ycjh-bgzxm0020 and Chongqing Natural Science Foundation under Grant No. CSTB2024NSCQ-JQX0022. R.G.C. is supported by the National Key Research and Development Program of China Grants No. 2020YFC2201502 and No. 2021YFA0718304 and by National Natural Science Foundation of China Grants No. 12235019. The numerical calculations in this study were carried out on the ORISE Supercomputer.

 \appendix

\section{Equations of Motion on Lattice}\label{App:EOM}
The Higgs field $\Phi$ is assigned to lattice sites, while gauge fields reside on links between neighboring sites via link variables:
\begin{align}
    U_i(t,x) &= 
    \exp\left[-\ii g\Delta x\frac{\sigma^a}{2} W_i^a(t,x)\right] \;,\\
    V_i(t,x) &= 
    \exp\left[-\ii g'\Delta x \,\frac12Y_i(t,x)\right] \;,\\
    V^\mathrm{ex}_i(t,x) &= 
    \exp\left[-\ii g'\Delta x \,\frac12Y^\mathrm{ex}_i(t,x)\right]\;.\label{eq:Vex}
\end{align}
Fields $\Phi$, $U_i$, $V_i$, and $V^\mathrm{ex}_i$ are evaluated at integer time steps $t, t+\Delta t,\ldots$, with their conjugate momenta $\Pi,\ E,\ F$ defined at midpoints $t+\Delta t/2, t+3\Delta t/2,\ldots$. 
Here $x+i$ ($x-i$) indicates displacement forward (backward) along the $i$-axis. 
The link variables $U_i(t,x)$, $V_i(t,x)$, $V^\text{ex}_i(t,x)$ parallelly transport $\Phi$ from site $x+i$ to $x$, while their Hermitian conjugates map $\Phi$ from $x$ to $x+i$. 
The covariant derivatives of $\Phi(t,x)$ are then:
\begin{align}
	D_i \Phi &= \frac{1}{\Delta x} \big[ U_i(t,x) V_i(t,x) V^{\mathrm{ex}}_{i}(t,x) \Phi(t,x+i) - \Phi(t,x)\big]\;, \\
	D_0 \Phi &= \frac{1}{\Delta t}\big[ \Phi(t+\Delta t,x) -\Phi(t,x) \big]\;.
\end{align}
Spatial plaquette operators are given by:
\begin{align}
    U_{ij}(t,x) &= U_j(t,x) U_i(t,x+j) U_j^\dagger (t,x+i) U_i^\dagger (t,x)\;, \\
    V_{ij}(t,x) &= V_j(t,x) V_i(t,x+j) V_j^\dagger (t,x+i) V_i^\dagger (t,x) \;,\\
    V^{\mathrm{ex}}_{ij}(t,x) &= V^{\mathrm{ex}}_{j}(t,x) V^{\mathrm{ex}}_{i}(t,x+j) V^{\mathrm{ex}\dagger}_{j} (t,x+i) V^{\mathrm{ex}\dagger}_{i}(t,x)\;.
\end{align}
Temporal-spatial plaquettes take the form:
\begin{align}
    U_{0i}(t,x) &=   U_i(t,x)U_i^\dagger(t+\Delta t,x)\;, \\
    V_{0i}(t,x) &= V_i(t,x)V_i^\dagger(t+\Delta t,x)\;, \\
    V^{\mathrm{ex}}_{0i}(t,x) &= V^{\mathrm{ex}}_{i}(t,x)V^{\mathrm{ex}\dagger}_{i}(t+\Delta t,x)\;.
\end{align}

The conjugate momenta relate to field updates via:
\begin{align}
\Phi(t+\Delta t,x) &= \Phi(t,x) +\Delta t \Pi(t+\Delta t/2,x) \;,\\
V_i(t+\Delta t,x) &= \frac{1}{2} g'\Delta x \Delta t E_i(t+\Delta t/2,x) V_i(t,x)\;, \\
U_i(t+\Delta t,x) &= g\Delta x \Delta t F_i(t+\Delta t/2,x) U_i(t,x)\;, 
\end{align}
Note that $V_{\mathrm{ex}i}(t,x)$ is static and thus lacks a conjugate momentum.

The lattice action reads:
\begin{widetext}
\begin{eqnarray}
    \nonumber
    S &=& \sum_{x,t} \Delta t \Delta x^3 \Bigg\{ \big(D_0\Phi \big)^\dagger \big(D_0\Phi \big) 
    - \sum_i \big(D_i\Phi \big)^\dagger \big(D_i\Phi \big) 
    -\mathcal{V}(\Phi) + \Big(\frac{2}{g\Delta t \Delta x}\Big)^2\sum_i \Big(1-\frac{1}{2}\text{Tr}~U_{0i}\Big) \\ 
    &&+\Big(\frac{2}{g'\Delta t \Delta x}\Big)^2 \sum_i \Big(1-\text{Re} ~V_{0i}\Big) 
    -\frac{2}{g^2 \Delta x^4} \sum_{i,j} \Big(1-\frac{1}{2} \text{Tr}~U_{ij}\Big) 
        - \frac{2}{g'^2\Delta x^4} \sum_{i,j} \Big(1-\text{Re}~V_{ij}\Big) \Bigg\} \notag \\
    &&+\Big(\frac{2}{g'\Delta t \Delta x}\Big)^2 \sum_i \Big(\mathrm{Im}~V_{0i}\Big) \Big(\mathrm{Im}~V_{\mathrm{ex}0i}\Big) 
    -\frac{2}{g'^2\Delta x^4} \sum_{i,j} \Big(\mathrm{Im}~V_{ij}\Big) \Big(\mathrm{Im}~V_{\mathrm{ex}ij}\Big).
    \label{actionlattice}
\end{eqnarray}
Setting the functional derivative of $S$ to zero yields equations of motion:
\begin{align}
    \Pi(t+\Delta t/2,x) =&\ \Pi(t-\Delta t/2,x)+\Delta t \Big\{ \frac{1}{\Delta x^2}\sum_i \big[ U_i(t,x)V_i(t,x)V_{\mathrm{ex}i}(t,x)\Phi(t,x+i)\;, \notag \\
	&-2\Phi(t,x)+U_i^\dagger(t,x-i) V_i^\dagger(t,x-i) V_{\mathrm{ex}i}^\dagger(t,x-i)\Phi(t,x-i)\big] - \frac{\partial \mathcal{V}}{\partial \Phi^\dagger}\Big\}\;,
    \label{eom-pi}
        \end{align}
        \begin{align}
    \text{Im} [E_k(t+\Delta t/2,x)] =&\ \text{Im} [E_k(t-\Delta t/2,x)]+\Delta t \Big\{ \frac{g'}{\Delta x}\text{Im}[\Phi^\dagger(t,x+k)U^\dagger_k(t,x) V^\dagger_k(t,x)V_{\mathrm{ex}k}^\dagger(t,x) \Phi(t,x)] \notag \\
    &\ -\frac{2}{g'\Delta x^3}\sum_i \text{Im} [V_k(t,x) V_i(t,x+k) V_k^\dagger (t,x+i) V_i^\dagger (t,x) \notag \\ 
    &\ + V_i(t,x-i) V_k(t,x) V_i^\dagger(t,x+k-i)V_k^\dagger(t,x-i)] \Big\} \;,\\
    \text{Tr} [i\sigma^m F_k(t+\Delta t/2,x)] =&\ \text{Tr} [i\sigma^m F_k(t-\Delta t/2,x)] +\Delta t\Big\{ \frac{g}{\Delta x} \text{Re} [\Phi^\dagger(t,x+k)U_k^\dagger(t,x)V_k^\dagger(t,x)V_{\mathrm{ex}k}^\dagger(t,x) i \sigma^m \Phi(t,x)] \notag \\
    &\ -\frac{1}{g\Delta x^3}\sum_i \text{Tr} [ i\sigma^m U_k(t,x) U_i(t,x+k) U_k^\dagger(t,x+i) U_i^\dagger(t,x) \notag\\
    &\ +i\sigma^m U_k(t,x) U_i^\dagger(t,x+k-i) U_k^\dagger(t,x-i) U_i(t,x-i) ]
		\Big\}\;.
\end{align}
Gauss constraints derived similarly are:
\begin{align}
    &\frac{1}{\Delta x}\sum_i
        \mathrm{Im}\big[E_i(t,x)-E_i(t,x-i)\big]
        -g'\mathrm{Im}\big[
            \Pi^\dagger(t,x)\Phi(t,x)\big] = 0 \;,\label{GaussU1} \\
    &\frac{1}{\Delta x}\mathrm{Tr}
            \sum_i\ii\sigma^m\big[F_i(t,x)-U^\dagger_i(t,x-i)F_i(t,x-i)U_i(t,x-i)\big]
            -g\mathrm{Re}\big[
                \Pi^\dagger(t,x)\ii\sigma^m\Phi(t,x)\big] = 0 \;.\label{GaussSU2}
\end{align}
\end{widetext}

\section{Initialization of gauge fields}\label{App:init}
The gauge fields are initialized to zero values, resulting in link variables of $1$ for U(1) and $1_{2\times2}$ for SU(2). To satisfy the Gauss constraint (Eq.~\eqref{gauss2}), the conjugate momentum fields of the gauge fields require initialization through specific procedures.

We first define
\begin{align}
    J_0^Y(t,x) &= \mathrm{Im}[\Pi^\dagger(t,x)\Phi(t,x)],\\
    J_0^a(t,x) &= \mathrm{Re}[\Pi^\dagger(t,x)\ii\sigma^a\Phi(t,x)].
\end{align}
Given $V_i(t=0,x)=1$ and $U_i(t=0,x)=1_{2\times2}$, the Gauss constraints in Eqs.~\eqref{GaussU1} and \eqref{GaussSU2} simplify to
\begin{align}
    &\frac{1}{\Delta x}\sum_i
        \mathrm{Im}\big[E_i(x)-E_i(x-i)\big] = g'J_0^Y(x), \label{U1gaussinit}\\
    &\frac{1}{\Delta x}\mathrm{Tr}
            \sum_i\ii\sigma^m\big[F_i(x)-F_i(x-i)\big] = gJ_0^a(x).\label{SU2gaussinit}
\end{align}
The $t=0$ specification is omitted here for conciseness. We introduce $\mathcal{\dot Y}_i(k)$ and $\mathcal{\dot W}_i^a(k)$ as the Fourier transforms of $\mathrm{Im}~E_i(x)$ and $\mathrm{Tr}[\ii\sigma^mF_i(x)]$ respectively:
\begin{align}
     \mathrm{Im}[E_i(x)]&= \frac{1}{N^3}\sum_k\mathcal{\dot Y}_i(k)e^{-\ii2\pi k(x+i/2)/N},\label{FourierE}\\
     F_i^m(x)=\mathrm{Tr}[\ii\sigma^mF_i(x)]&= \frac{1}{N^3}\sum_x\mathcal{\dot W}_i^m(k)e^{-\ii2\pi k(x+i/2)/N}.\label{FourierF}
\end{align}
These definitions utilize the connections between gauge momentum fields and link variables:
\begin{align}
    E_i(x) &= \frac{2}{g'\Delta x\Delta t}\exp[-\ii g'\Delta x\, \frac12\dot Y_i(x)], \\
    F_i(x) &= \frac{1}{g\Delta x\Delta t}\exp[-\ii g\Delta x \frac{\sigma^a}{2}\dot W^a_i(x)].
\end{align} 
Fourier transforming both sides of Eqs.~\eqref{U1gaussinit} and \eqref{SU2gaussinit}, while accounting for link fields spanning lattice sites, yields:
\begin{widetext}
\begin{align}
    &\ \frac{1}{\Delta xN^3}\sum_x\sum_i\sum_k[\mathcal{\dot Y}_i(k)e^{-\ii2\pi k(x+i/2)/N}-\mathcal{\dot Y}_i(k)e^{-\ii2\pi k(\mathbf{x-\hat i}/2)/N}]e^{\ii2\pi k'x/N}\notag\\
    =&\ \frac{1}{\Delta xN^3}\sum_k\sum_i\sum_x\mathcal{\dot Y}_i(k)e^{-\ii2\pi(k-k')x/N}[e^{-\ii\pi k_i/N}-e^{\ii\pi k_i/N}] \notag\\
    =&\ \frac{-2\ii}{\Delta x}\sum_i\mathcal{\dot Y}_i(k')\sin\frac{\pi k'_i}{N}\notag \\
    =&\ g'\sum_xJ_0^Y(x)e^{\ii2\pi k'x/N}
    =g'J_0^Y(k').
\end{align}
\end{widetext}
Solving for $\mathcal{\dot Y}_i(k)$ gives:
\begin{align}
    \mathcal{\dot Y}_i(k) = \sin\frac{\pi k_i}{N}\left(\sum_i\sin^2\frac{\pi k_i}{N}\right)^{-1}\frac\ii2g'\Delta xJ_0^Y(k).
\end{align}
Analogous derivation for the SU(2) sector produces:
\begin{align}
    \mathcal{\dot W}^m_i(k) = \sin\frac{\pi k_i}{N}\left(\sum_i\sin^2\frac{\pi k_i}{N}\right)^{-1}\frac\ii2g\Delta xJ_0^m(k).
\end{align}
The coordinate-space conjugate momenta follow via inverse Fourier transforms using Eqs.~\eqref{FourierE} and \eqref{FourierF}. Complete link variables are then initialized through:
\begin{align}
    E_i(x) =&\  \sqrt{\left(\frac{2}{g'\Delta x\Delta t}\right)^2 - [\mathrm{Im}~E_i(x)]^2} + \ii \mathrm{Im}~E_i(x) \label{Ek}, \\
    F_i(x) =&\  \sqrt{\left(\frac{1}{g\Delta x\Delta t}\right)^2
    -\sum_m\left(\frac{F^m_i(x)}{2}\right)^2}
    \begin{pmatrix}
        1 & 0 \\
        0 & 1
    \end{pmatrix}
     \notag\\&\ - \ii \frac{\sigma^m}{2}F^m_i(x).
\end{align}

\section{Hypermagnetic field on lattice}
\label{sec:hyMFlat}
The infinite correlation length hypermagnetic field configuration with helicity takes the form $Y^{\mathrm{ex}\mu} = (0,\ 0,\ xB_Y^\mathrm{ex},\ h_\mathrm{factor}LB_Y^\mathrm{ex})$, satisfying
\begin{align}
    \nabla \times \boldsymbol{Y}^\mathrm{ex}
    = \begin{vmatrix}
        \hat x & \hat y & \hat z \\
        \partial_x & \partial_y & \partial_z \\
        0 & xB_Y^\mathrm{ex} & h_\mathrm{factor}LB_Y^\mathrm{ex}
    \end{vmatrix}
    = \partial_x(xB_Y^\mathrm{ex})\hat z = B_Y^\mathrm{ex}\hat z, 
\end{align}
where $B_Y^\mathrm{ex}$ denotes constant field strength, $h_\mathrm{factor} \in [-1,\ 1]$ adjusts helicity, and $L = N\Delta x$ is the lattice size. 

Discretization replaces continuum coordinates $(x,y,z)$ with lattice indices $(n_1,n_2,n_3)$:
\begin{align}
    \partial_x(xB_Y^\mathrm{ex})\hat z &= \frac{1}{\Delta x}[(n_1+1)\Delta x - n_1\Delta x]B_Y^\mathrm{ex}\hat z \\
    &= B_Y^\mathrm{ex}\hat z\;,\qquad n_1 \neq N-1.
\end{align}
Boundary behavior differs due to periodicity. At $n_1 = N-1$, $\boldsymbol{Y}^{\mathrm{ex}}(t,N,n_2,n_3) = \boldsymbol{Y}^{\mathrm{ex}}(t,0,y,z) = 0$ yields
\begin{align}
    \partial_x(xB_Y^\mathrm{ex})\hat z|_{x=(N-1)\Delta x} &= \frac{1}{\Delta x}[0 - (N-1)\Delta x]B_Y^\mathrm{ex}\hat z\notag\\
    &= -(N-1)B_Y^\mathrm{ex}\hat z \neq B_Y^\mathrm{ex}\hat z. 
\end{align}
This discrepancy is resolved by setting $Y_1^{\mathrm{ex}}(t,N-1,n_2,n_3)=-Nn_2\Delta x B_Y^\mathrm{ex}$, resulting in
\begin{align}
    &\ \nabla \times \boldsymbol{Y}^\mathrm{ex}|_{x=(N-1)\Delta x} \notag\\
    =&\ (\partial_x Y^\mathrm{ex}_2 - \partial_y Y^\mathrm{ex}_1)|_{x=(N-1)\Delta x}\hat z \notag \\
    =&\ \{(-(N-1)B_Y^\mathrm{ex} + \frac{N}{\Delta x}[(n_2+1)\Delta x-n_2\Delta x]B_Y^\mathrm{ex}\}\hat z \notag \\
    =&\ B_Y^\mathrm{ex}\hat z. 
\end{align}
ensuring boundary-internal consistency. The complete external U(1) field specification is:
\begin{align}
    Y_1^{\mathrm{ex}}(t,N-1,n_2,n_3)&=-Nn_2\Delta x B_Y^\mathrm{ex},\\
    Y_1^{\mathrm{ex}}(t,n_1\neq N-1,n_2,n_3)&=0,\\
    Y_2^{\mathrm{ex}}(t,n_1,n_2,n_3)&=n_1\Delta xB_Y^\mathrm{ex},\\
    Y_3^{\mathrm{ex}}(t,n_1,n_2,n_3) &= h_\mathrm{factor}LB_{Y}^\mathrm{ex},\\
    Y_0^{\mathrm{ex}}(t,n_1,n_2,n_3) = 0.
\end{align}
The quantization condition~\cite{Bali_2012} $B^\mathrm{ex}_Y (N\Delta x)^2 = \frac{2}{g'} 2\pi N_B \label{qc},\ N_B\in \mathbb{Z}$ is applied.

Finite correlation length hyperMFs utilize the spectral representation~\cite{Brandenburg_2020}:
\begin{align}
    \Tilde{B}_{Yi}^\mathrm{ex}(\boldsymbol{k}) = B_\mathrm{ini}\Theta(k-k_\mathrm{UV})\left(\delta_{ij}-\hat{k}_i\hat{k}_j-\ii\sigma_\mathrm{M}\varepsilon_{ijl}\hat{k}_l\right)g_j(\boldsymbol{k})k^n. \label{eq:specB}
\end{align}
The corresponding vector potential in Fourier space is
\begin{align}
    \Tilde{Y}^\mathrm{ex}_l(\boldsymbol{k}) = \ii B_\mathrm{ini}\left(\varepsilon_{lmn}\hat{k}_m-\ii\sigma_\mathrm{M}\delta_{ln}\right)g_n(\boldsymbol{k})k^{n-1}. \label{eq:specY}
\end{align}
Consistency with $\boldsymbol{B}^\mathrm{ex}(\boldsymbol{x})=\nabla\times\boldsymbol{Y}^\mathrm{ex}(\boldsymbol{x})$ follows from
\begin{align}
    \int \frac{\mathrm{d}^3\boldsymbol{k}}{(2\pi)^3}\,\Tilde{\boldsymbol{B}}^\mathrm{ex}(\boldsymbol{k})e^{\ii\boldsymbol{k\cdot x}} &= \nabla\times\int \frac{\mathrm{d}^3\boldsymbol{k}}{(2\pi)^3}\,\Tilde{\boldsymbol{Y}}^\mathrm{ex}(\boldsymbol{k})e^{\ii\boldsymbol{k\cdot x}} \notag\\
    &= \int \frac{\mathrm{d}^3\boldsymbol{k}}{(2\pi)^3}\,\ii\boldsymbol{k}\times\Tilde{\boldsymbol{Y}}^\mathrm{ex}(\boldsymbol{k})e^{\ii\boldsymbol{k\cdot x}}.
\end{align}
The relation $\Tilde{B}^\mathrm{ex}_i(\boldsymbol{k})=\ii\varepsilon_{ijl}k_j\Tilde{Y}^\mathrm{ex}_l(\boldsymbol{k})$ combined with \eqref{eq:specY} yields
\begin{align}
    &\ \ii\varepsilon_{ijl}k_j\Tilde{Y}^\mathrm{ex}_l(\boldsymbol{k}) \notag\\
    =&\ -\varepsilon_{ijl}k_j B_\mathrm{ini}\left(\varepsilon_{lmn}\hat{k}_m-\ii\sigma_\mathrm{M}\delta_{ln}\right)g_n(\boldsymbol{k})k^{n-1} \notag\\
    =&\  -B_\mathrm{ini}\left[(\delta_{im}\delta_{jn}-\delta_{in}\delta_{jm})\hat{k}_m\hat{k}_j-\ii\sigma_\mathrm{M}\varepsilon_{ijn}\hat{k}_j\right]g_n(\boldsymbol{k})k^{n} \notag\\
    =&\  B_\mathrm{ini}\left(-\hat{k}_i\hat{k}_n+\delta_{in}+\ii\sigma_\mathrm{M}\varepsilon_{ijn}\hat{k}_j\right)g_n(\boldsymbol{k})k^{n} \notag\\
    =&\  B_\mathrm{ini}\left(\delta_{ij}-\hat{k}_i\hat{k}_j-\ii\sigma_\mathrm{M}\varepsilon_{ijl}\hat{k}_l\right)g_j(\boldsymbol{k})k^{n},
\end{align}
matching \eqref{eq:specB}. 

Implementation of \eqref{eq:specB} proceeds as:
\begin{itemize}
    \item Generate coordinate-space Gaussian random field $\boldsymbol{g}(\boldsymbol{x})$ with $\delta$-correlation
    \item Fourier transform to obtain $\boldsymbol{g}(\boldsymbol{k}) = \int\mathrm{d}^3\boldsymbol{x}\, \boldsymbol{g}(\boldsymbol{x})e^{-\ii\boldsymbol{k\cdot x}}$
    \item Compute $\boldsymbol{\Tilde{Y}}^\mathrm{ex}(\boldsymbol{k})$ via \eqref{eq:specY}
    \item Apply inverse Fourier transform: $\boldsymbol{Y}^\mathrm{ex}(\boldsymbol{x}) = \int \frac{\mathrm{d}^3\boldsymbol{k}}{(2\pi)^3}\Tilde{\boldsymbol{Y}}^\mathrm{ex}(\boldsymbol{k})e^{\ii\boldsymbol{k\cdot x}}$
    \item Construct link variables $V^\mathrm{ex}_i$ using \eqref{eq:Vex}
\end{itemize}

	\bibliography{reference}

\end{document}